\documentclass[journal]{IEEEtran}

\usepackage{amsmath,graphicx}
\usepackage{mathrsfs}
\usepackage{amsfonts}
\usepackage[linesnumbered,ruled,vlined]{algorithm2e}
\usepackage{mathtools}
\usepackage{commath}
\usepackage{dblfloatfix}
\usepackage{float}
\usepackage{afterpage}
\usepackage{amsthm}
\usepackage{color}

\DeclareMathAlphabet\mathbfcal{OMS}{cmsy}{b}{n}

\begin{document}

\title{Third-Order Statistics Reconstruction from Compressive Measurements}

\author{Yanbo Wang,~\IEEEmembership{Student Member,~IEEE}
        and Zhi Tian,~\IEEEmembership{Fellow,~IEEE}
\thanks{This work was supported in part by the National Science Foundation Grant \#CCF-1527396. Parts of this paper were presented at the IEEE International Conference on Acoustics, Speech, and Signal Processing, Barcelona, Spain, May, 2020 \cite{wang2020cumulant} and at the Asilomar Conference on Signals, Systems, and Computers, Pacific Grove, CA, USA, November, 2020 \cite{Wang2020third}.

The authors are with the Department of Electrical and Computer Engineering, George Mason University, Fairfax, VA 22030 USA (e-mail: ywang80@gmu.edu, ztian1@gmu.edu).}
}

\markboth{IEEE TRANSACTIONS ON SIGNAL PROCESSING,~Vol.~\#, No.~\#, July~2020}%
{Shell \MakeLowercase{\textit{et al.}}: Bare Demo of IEEEtran.cls for IEEE Journals}

\maketitle

\begin{abstract}
Estimation of third-order statistics relies on the availability of a huge amount of data records, which can pose severe challenges on the data collecting hardware in terms of considerable storage costs, overwhelming energy consumption, and unaffordably high sampling rate especially when dealing with high-dimensional data such as wideband signals. To overcome these challenges, this paper focuses on the reconstruction of the third-order cumulants under the compressive sensing framework. Specifically, this paper derives a transformed linear system that directly connects the cross-cumulants of compressive measurements to the desired third-order statistics. We provide sufficient conditions for lossless third-order cumulant reconstruction via solving simple least-squares, and quantify the strongest achievable compression ratio. To reduce the computational burden, we also propose an approach to recover diagonal cumulant slices directly from compressive measurements, which is useful when the cumulant slices are sufficient for the inference task at hand. As testified by extensive simulations, the developed joint sampling and reconstruction approaches to third-order statistics estimation are able to reduce the required sampling rates significantly by exploiting the cumulant structure resulting from signal stationarity, even in the absence of any sparsity constraints on the signal or cumulants.


\end{abstract}


\IEEEpeerreviewmaketitle

\section{Introduction}

\IEEEPARstart{H}{igh}-dimensional signals are ubiquitously featured in contemporary systems, which provide rich information for enhancing statistical inference performance, but at the same time pose unprecedented challenges to traditional signal processing theories and methods. As the signal bandwidth increases significantly, the sampling rate for digital conversion ruled by the Shannon-Nyquist sampling theorem can be exceedingly high in the wideband regime and hence easily reach the limit of analog-to-digital converter (ADC) hardware which is a main bottleneck of many current signal processing systems. In order to alleviate the burden on ADCs, compressed sensing (CS) is shown to be able to accurately recover signals from sub-Nyquist-rate samples by capitalizing on signal sparsity in the spectrum or the edge spectrum \cite{donoho2006compressed} \cite{candes2006robust} \cite{tian2007compressed}. Closely related works can also be found in the so-called spectrum blind sampling (SBS) framework \cite{bresler2008spectrum} \cite{mishali2010theory} where analog-to-digital sub-Nyquist samplers are designed for spectrum reconstruction under the sparse spectrum assumption. 

Another line of compressive sampling research advocates direct second-order statistics (SOS) reconstruction from sub-Nyquist samples even in the absence of signal sparsity. In \cite{leus2011power} \cite{tian2011cyclic} \cite{tian2011cyclicf}, a compressive covariance sensing (CCS) framework is proposed to recover sparse cyclic spectrum from compressive measurements for cyclostationary processes. Taking stationary signals as a special case therein, it shows that lossless power spectrum recovery is possible even for non-sparse signals by capitalizing on the Toeplitz structure of covariances. For improved reconstruction accuracy, generalizations are considered in \cite{leus2011recovering} \cite{ariananda2012compressive}, where all significant lags of cross-correlation functions among the compressive outputs are exploited to induce an overdetermined system for non-sparse cyclic spectrum. In \cite{ariananda2012compressive} \cite{romero2016compressive} \cite{ariananda2011multi}, sampler design for CCS is formulated into a minimal sparse ruler problem with the goal of achieving the strongest compression. Other 
 sparse sampler designs with closed-form expressions can be found in \cite{vaidyanathan2010sparse} \cite{pal2011coprime}, in which two uniform sub-Nyquist-rate samplers are exploited for sinusoids frequency estimation based on the co-array concept and properties of coprime numbers.



For non-Gaussian signals appeared in many real-world applications, higher-order statistics (HOS) preserve non-trivial information that are absent in SOS \cite{mendel1991tutorial} \cite{nikias1993signal} \cite{giannakis1990nonparametric}. Such information can be used in many tasks that cannot be accomplished using SOS only, such as suppressing additive colored Gaussian noise, blindly identifying non-minimum phase systems, to name just a few. In practice, accurate HOS estimation hinges on the availability of a huge amount of data records, which poses challenges for data acquisition. Besides the exceedingly high cost in data storage, the sampler consumes substantial energy in order to collect a huge amount of data over a long sampling time, which is unaffordable for instance in power-hungry sensor networks. These challenges become even pronounced when dealing with wideband signals because high sampling rate requirements not only overburden ADCs but also consume a large amount of power. In view of these challenges, it is well motivated to develop compressive sampling techniques for HOS estimation.

Despite of the well-documented merits of HOS in various signal processing tasks, the literature on HOS estimation under the compressive sampling framework is scant. In \cite{cao2013compressive}, third-order cumulants of compressive measurements are calculated and directly used for spectrum sensing tasks, but it does not concern the reconstruction of uncompressed third-order cumulants.  In \cite{lim2015compressive}, estimating higher-order cyclostationary features from compressive samples is studied by taking into account of sparsity in the frequency domain. However, \cite{lim2015compressive} recovers only a part of cumulant values instead of the entire cumulant tensor and does not generalize to cases in the absence of sparsity assumption. To reduce the computational complexity, the generalization of co-prime sampling for HOS estimation is studied in \cite{wu2015higher}. In \cite{sun2018sparse}, sparse representation of HOS is explored. However, the sampler design, the HOS recovery guarantees and the achievable sampling rate reduction are not considered in \cite{wu2015higher} and \cite{sun2018sparse}.



In this paper, we put forward a systematic and simple approach to the estimation of third-order statistics under the compressive sampling framework with recovery guarantees. Different from existing compressed sensing techniques for recovering one-dimensional deterministic signals, the collected compressive samples do not have a direct linear relationship with the third-order statistics to be recovered. To circumvent this difficulty, this paper derives a transformed linear system that connects the cross-cumulants of compressive outputs to the desired third-order statistics. Theoretically, the proposed approach can losslessly recover the unknown third-order statistics of a wide-sense stationary signal via solving simple least-squares (LS). We analyze sufficient conditions for lossless third-order statistics reconstruction by using the proposed approach, and derive the strongest achievable compression ratio. Realizing that the diagonal cumulant slice is sufficient for many signal processing tasks, we also propose an approach to reconstruct arbitrary-order cumulant slices directly from compressive measurements without having to recover the entire cumulant tensor, which interestingly subsumes CCS as a special case. Extensive simulations demonstrate that the proposed approaches are able to reduce the required sampling rate significantly without imposing sparsity constraints on the signal or its cumulants.


The rest of the paper is organized as follows. Section II reviews some key definitions on third-order statistics. Section III gives the signal model and problem statement. Section IV describes the proposed compressive cumulant estimation approaches.
Section V considers the symmetry property of third-order cumulants to further enhance the estimation performance. Section VI derives a compressive cumulant slice sensing approach to recover the diagonal cumulant slice directly. Corroborating simulation results are provided in Section VII, followed by a concluding summary in Section VIII.

\section{Preliminaries}

This section introduces relevant notation and reviews some basic definitions of third-order statistics.

\subsection{Notation}
We use lower case boldface letters, upper case boldface letters, and calligraphic boldface letters, such as $\mathbf{c}$, $\mathbf{C}$, $\mathbfcal{C}$, to denote vectors, matrices and tensors respectively. The operator $\text{vec}(\cdot)$ stacks all columns (slices) of a matrix (tensor) into a large column vector. Calligraphic letters such as $\mathbb{K}$ are used to denote sets, and $|\mathbb{K}|$ represents the corresponding cardinality.  We use $\star$ to denote the convolution operator, $\circ$ the outer product, and $\otimes$ the Kronecker product.  Three-fold Kronecker product of vector $\mathbf{y}$ is denoted as $\mathbf{y}^{(3)}=\mathbf{y}\otimes\mathbf{y}\otimes\mathbf{y}$.  We use $(\cdot)^T$ to represent transposition and $(\cdot)^\dag$ the Moore-Penrose pseudo inverse. We write the $\ell_2$-norm of a vector as $||\cdot||_2$. We use $\mathbf{I}_N$ to denote the identity matrix of size $N\times N$.

\subsection{Third-Order Statistics}

This subsection reviews some basic definitions that are used in the ensuing sections \cite{nikias1993signal}.

\emph{Definition 1:} For zero-mean real random variables $x_1, x_2, x_3$, the third-order cumulant is given by:
\begin{equation}
cum(x_1, x_2, x_3)= E\{x_1x_2x_3\}.
\end{equation}

\emph{Definition 2:} Let $\{x(n)\}$ be a \emph{wide-sense stationary random process}. The third-order cumulant of this process, denoted as $c_{3,x}(\tau_1, \tau_2)$, is defined as the joint third-order cumulant of random variables $x(n), x(n+\tau_1),x(n+\tau_2)$, i.e., 
\begin{equation}
c_{3,x}(\tau_1, \tau_2)=cum(x(n), x(n+\tau_1), x(n+\tau_2)). \label{3 stationary cumulants}
\end{equation}
Because $x(n)$ is \emph{stationary}, its cumulant is dependent only on time-lags $\tau_1$ and $\tau_2$ while independent on the time origin $n$.


\emph{Definition 3:} Let $\{x(n)\}$ be a \emph{non-stationary random process}. Its third-order cumulant is defined as
\begin{equation}
c_{3,x}(n; \tau_1, \tau_2)=cum(x(n), x(n+\tau_1), x(n+\tau_2)), \label{3 nonstationary cumulants}
\end{equation}
which is dependent on both time-lags $\tau_1$ and $\tau_2$ and the time origin $n$ because of the nonstationarity.

\emph{Definition 4:} Let $\mathbf{y}(k)$ be a vector random process of dimension $M$, i.e., $\mathbf{y}=[y_1(k), y_2(k), \cdots, y_M(k)]^T$. The third-order cross-cumulant of elements $y_{i_1}(k), y_{i_2}(k+\tau_1), y_{i_3}(k+\tau_2), \forall i_1, i_2, i_3 = 1, 2, \cdots M$ in $\mathbf{y}(k)$ is defined as 
\begin{equation}
c_{y_{i_1}, y_{i_2}, y_{i_3}}(\tau_1, \tau_2) = E\{y_{i_1}(k)y_{i_2}(k+\tau_1)y_{i_2}(k+\tau_2) \}.
\end{equation}
\emph{Definition 5:}  Let $x(n)$ be a zero-mean $q$th-order stationary process. The $q$th-order cumulant of this process, denoted by $c_{q, x}(\tau_1,\tau_2,...,\tau_{q-1})$, is defined as the joint $q$th-order cumulant of random variables $x(n), x(n+\tau_1), ..., y(n+\tau_{q-1})$, 
\begin{equation}
\begin{aligned}
c_{q, x}&(\tau_1, \tau_2 ...,\tau_{q-1})=\\
&cum(x(n), x(n+\tau_1), ..., x(n+\tau_{q-1})). \label{kth}
\end{aligned}
\end{equation}

\section{System Model and Problem Statement}

\begin{figure}
  \centering
  \centerline{\includegraphics[width=7cm, height=3.5cm]{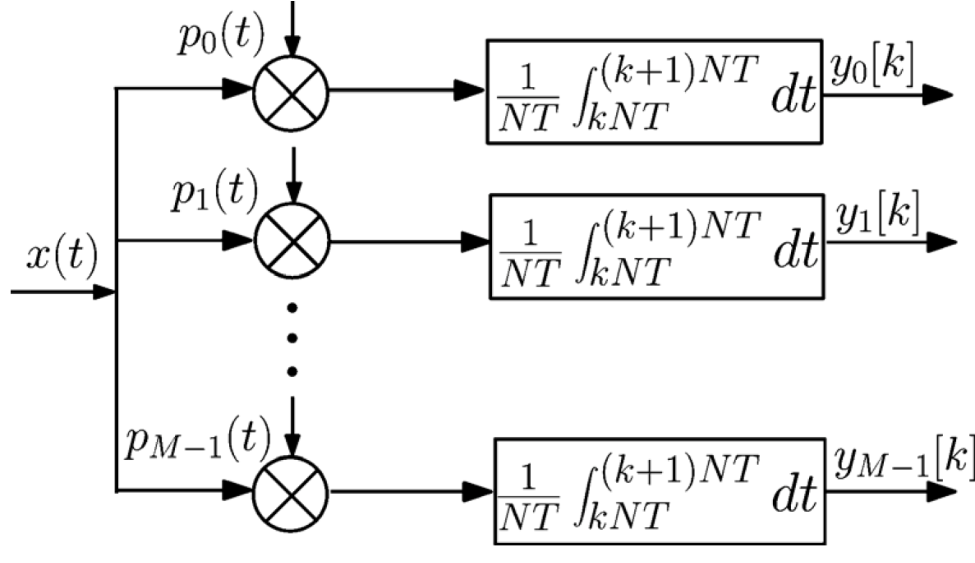}}
  \caption{\label{figure_label1} Illustration of AIC implementation \cite{ariananda2012compressive}. } 
\end{figure}

The third- and higher-order cumulants of real-valued stationary processes can be useful in many real-world applications such as real harmonics retrieval in the additive colored noise setting, quadratic phase coupling detection in nonlinear processes, and time delay estimation for source bearing and range calculation \cite{mendel1991tutorial} \cite{nikias1993signal}. Consider w.l.o.g. a real-valued wide-sense stationary analog signal $x(t)$ which is bandlimited with bandwidth $\frac{1}{2T}$. For digital processing at a sub-Nyquist sampling rate, an analog to information converter (AIC) can be adopted \cite{ariananda2012compressive}, as shown in Fig. \ref{figure_label1}. Equipped with $M$ branches, the adopted sampler first modulates the signal $x(t)$ with a periodic waveform $p_i(t)$ of period $NT$ in the $i$th branch, and then convert the modulated signal in each branch to the digital version via an integrate-and-dump device operating with period $NT$. Through the AIC, a total of $M$ samples are collected for every $NT$ seconds with $M < N$, resulting in an effective sampling rate of $M/NT$ that is $M/N$ times the Nyquist rate. The compression ratio $M/N$ determines the reduced sampling rate. 
Evidently, the output of the $i$th branch at the $k$th sampling instant is given by
\begin{equation}
\begin{aligned}
y_i[k] &= \frac{1}{NT}\int_{kNT}^{(k+1)NT} p_i(t)x(t)dt\\
&=\frac{1}{T}\int_{kNT}^{(k+1)NT} e_i(t-kNT)x(t)dt  
\end{aligned} \label{sampling device}
\end{equation}
where $e_i(t):=\frac{1}{N}p_i(t)$ for $0\leq t<NT$ and $e_i(t):=0$ otherwise. By assuming that $e_i(t)$ is a piecewise constant function having constant values in every interval of length $T$, i.e., for $e_i(t)=e_i[-n]$ for $nT\leq t<(n+1)T$, where $n=0,1,...,N-1$, the output (\ref{sampling device}) can be rewritten as
\begin{equation}
\begin{aligned}
y_i[k] &= \sum_{n=0}^{N-1}e_i[-n]\frac{1}{T}\int_{(kN+n)T}^{(kN+n+1)T}x(t)dt\\
&=\sum_{n=0}^{N-1}e_i[-n]x[kN+n]\\
&=\sum_{n=1-N}^{0}e_i[n]x[kN-n]
\end{aligned} \label{sampling device rewritten}
\end{equation}
where $x[n]=\frac{1}{T}\int_{nT}^{(n+1)T}x(t)dt$ is the Nyquist digital version of $x(t)$, which is not acquired in practical implementation for sake of reducing the sampling rate. The choice of $e_i(t)$ depends on the sampler design.  Random Gaussian sampler, similar to \cite{laska2007theory}, is adopted in this paper, which will be detailed in the ensuing sections.

The sampling strategy in (\ref{sampling device rewritten}) can be alternatively interpreted as a two-step operation. It first convolves $x[n]$ with a length-$N$ digital filter $e_i[n]$ and then feeds the outcome to an $N$-fold decimation operator to obtain the compressive outputs, i.e., $y_i[k]=z_i[kN]$, where
\begin{equation}
z_i[n] = e_i[n] \star x[n]= \sum_{m=1-N}^0 e_i[m]x[n-m] \label{digital z}.
\end{equation}
It is worth noting that our system model can adopt any linear compressive sampler besides the AIC, for which (\ref{sampling device rewritten}) and (\ref{digital z}) hold for some $e_i[n]$. The relationship $y_i[k] = z_i[kN]$ is useful, which is illustrated via the AIC. 

With the above model in mind, the overarching goal of this paper is to accurately reconstruct the third-order cumulant $c_{3,x}( \tau_1, \tau_2)=E\{x[n]x[n+\tau_1]x[n+\tau_2]\}$ of $x[n]$ based on the obtained sub-Nyquist-rate samples $\left\lbrace y_i[k]\right\rbrace_{i,k}$. The main contribution of this work is that all the $M^3$ different cross-cumulants $c_{y_{i_1}, y_{i_2}, y_{i_3}}( \tau_1, \tau_2)=E\{y_{i_1}[k] y_{i_2}[k+\tau_1],y_{i_3}[k+\tau_2]\}$ for $i_1, i_2, i_3= 0, 1, \cdots, M-1$ are leveraged to recover $O(N^2)$ unknown variables $c_{3,x}( \tau_1, \tau_2)$ for $\tau_1, \tau_2=0,1, \cdots, N-1$, which makes the reconstruction feasible from sub-Nyquist-rate samples even in the absence of sparsity constraints on $x[n]$ or its cumulants. Attractively, such a reconstruction strategy can be formulated as an overdetermined linear system as analyzed next, leading to a unique solution via solving simple LS. 

\section{Third-Order Cumulants Reconstruction}
\label{sec:majhead}

In this section, we develop two compressive third-order cumulants sensing (C3CS) approaches.

\subsection{The direct C3CS Approach}
\label{ssec:subhead}

This subsection presents a direct C3CS approach for $c_{3,x}(\tau_1, \tau_2)$ reconstruction from $c_{y_{i_1}, y_{i_2}, y_{i_3}}(\tau_1, \tau_2)$ for all $i_1, i_2, i_3= 0,1,...,M-1$. Considering that $y_i[k]=z_i[kN]$, the third-order cross-cumulant of sub-Nyquist-rate samples $y_{i_1}[n], y_{i_2}[n], y_{i_3}[n]$ and that of Nyquist-rate samples $z_{i_1}[n], z_{i_2}[n], z_{i_3}[n]$ are related through:
\begin{equation}
\begin{aligned}
&c_{y_{i_1}, y_{i_2}, y_{i_3}}(\tau_1, \tau_2) \\
&= E\{y_{i_1}[k]y_{i_2}[k+\tau_1] y_{i_3}[k+\tau_2]\}\\
&=E\{z_{i_1}[kN]z_{i_2}[(k+\tau_1)N]z_{i_3}[(k+\tau_2)N\}\\
&= c_{z_{i_1}, z_{i_2}, z_{i_3}}( \tau_1 N, \tau_2 N).
\end{aligned} \label{cy and cz}
\end{equation}

Viewing $z_i[n]$ as the convolution of $e_i[n]$ and $x[n]$, $c_{z_{i_1}, z_{i_2}, z_{i_3}}(\tau_1, \tau_2)$ can be obtained from a two-dimensional linear convolution \cite{mendel1991tutorial}
\begin{equation}
\begin{aligned}
&c_{z_{i_1}, z_{i_2},z_{i_3}}(\tau_1 , \tau_2)\\
&=\sum_{m_1=1-N}^{N-1}\sum_{m_2=1-N}^{N-1} c_{e_{i_1}, e_{i_2}, e_{i_3}}(m_1, m_2)\\
&\qquad \qquad \qquad \qquad \qquad \qquad c_{3,x}( \tau_1-m_1, \tau_2-m_2)
\end{aligned} \label{cz and cx}
\end{equation}
where $c_{e_{i_1}, e_{i_2}, e_{i_3}}(m_1, m_2)$ is the “deterministic” third-order cross-cumulant of $e_{i_1}[n]$,$e_{i_2}[n]$ and $e_{i_3}[n]$ in the form
\begin{equation}
\begin{aligned}
c_{e_{i_1}, e_{i_2}, e_{i_3}}(m_1, m_2) = \sum_{n=1-N}^0 e_{i_1}[n] e_{i_2}[n+m_1]e_{i_3}[n+m_2]. \label{c3ecumu}
\end{aligned}
\end{equation}

Combining (\ref{cy and cz}) and (\ref{cz and cx}) yields
\begin{equation}
\begin{aligned}
&c_{y_{i_1}, y_{i_2}, y_{i_3}}(\tau_1, \tau_2)= c_{z_{i_1}, z_{i_2}, z_{i_3}}(\tau_1 N, \tau_2 N)&\\
&=\sum_{m_1=1-N}^{N-1}\sum_{m_2=1-N}^{N-1}c_{e_{i_1}, e_{i_2},e_{i_3}}(m_1, m_2)\\
& \qquad \qquad \qquad \qquad \qquad c_{3,x}( \tau_1 N-m_1, \tau_2 N-m_2) & \\
&=\sum_{a=0}^1\sum_{b=0}^1  \mathbf{c}^T_{e_{i_1}, e_{i_2} e_{i_3}}[a,b]\mathbf{c}_{3,x}[\tau_1 -a, \tau_2 -b]&
\end{aligned} \label{vec convolution}
\end{equation}
where the concise vector representation in the last equality is based on the following vector definitions:
\begin{equation}
\mathbf{c}_{3, x}[\tau_1, \tau_2]=\text{vec}(\mathbf{C}_{3,x}[\tau_1, \tau_2]^T), \label{c3x}
\end{equation}
\begin{equation}
\mathbf{c}_{e_{i_1}, e_{i_2}, e_{i_3}}[a, b]=\text{vec}(\mathbf{C}_{e_{i_1}, e_{i_2}, e_{i_3}}[a,b]^T), \label{c3e}
\end{equation}
with matrices $\mathbf{C}_{3,x}[\tau_1, \tau_2]$ and $\mathbf{C}_{e_{i_1}, e_{i_2}, e_{i_3}}[a,b]$ given by (\ref{C3x}) and (\ref{Ce1e2e3}) respectively.
\begin{table*}[ht]
\begin{equation}
\begin{aligned}
\mathbf{C}_{3,x}[\tau_1, \tau_2]=\left[
  \begin{array}{ccc} 
    c_{3,x}(\tau_1N, \tau_2N) & \cdots& c_{3,x}(\tau_1N, \tau_2N+(N-1))\\ 
   \vdots & \vdots & \vdots  \\ 
    c_{3,x}(\tau_1N+(N-1), \tau_2N) & \cdots &  c_{3,x}(\tau_1N+(N-1), \tau_2N+(N-1))\\ 
  \end{array} 
\right]
\end{aligned} \label{C3x}
\end{equation}
\end{table*} 
\begin{table*}[ht]
\begin{equation}
\begin{aligned}
\mathbf{C}_{e_{i_1}, e_{i_2}, e_{i_3}}[a,b]=\left[
  \begin{array}{ccc} 
    c_{e_{i_1}, e_{i_2}, e_{i_3}}(aN, bN) & \cdots& c_{e_{i_1}, e_{i_2}, e_{i_3}}(aN, bN-(N-1))\\ 
   \vdots & \vdots & \vdots  \\ 
    c_{e_{i_1}, e_{i_2}, e_{i_3}}(aN-(N-1), bN) & \cdots &  c_{e_{i_1}, e_{i_2}, e_{i_3}}(aN-(N-1), bN-(N-1))\\ 
  \end{array} 
\right]
\end{aligned} \label{Ce1e2e3}
\end{equation}
\end{table*} 

We next stack all the $M^3$ cross-cumulant values $c_{y_{i_1}, y_{i_2}, y_{i_3}}(\tau_1, \tau_2)$ for $i_1, i_2, i_3= 0, 1, \cdots, M-1$ into a $M^3\times 1$ vector $\mathbf{c}_{3,y}[\tau_1, \tau_2]=[..., c_{y_{i_1}, y_{i_2}, y_{i_3}}( \tau_1, \tau_2) , ...]^T$ which according to (\ref{vec convolution}) can be expressed as
\begin{equation}
\begin{aligned}
&\mathbf{c}_{3,y}[\tau_1, \tau_2]
&=\sum_{a=0}^1\sum_{b=0}^1 \mathbf{C}_{3,e}[a, b] \mathbf{c}_{3,x}[ \tau_1 -a, \tau_2 -b]
\end{aligned} \label{matrix conv}
\end{equation}
where matrices $\mathbf{C}_{3,e}[a, b]:=\left[\dots, \mathbf{c}_{e_{i_1}, e_{i_2}, e_{i_3}}[a, b], \dots\right]^T$ (for $i_1, i_2, i_3= 0, 1, \cdots, M-1$ and $a,b\in \left\lbrace0,1\right\rbrace$) are of size $M^3\times N^2$ .

Although the bandlimitedness assumption on $x[n]$ in the frequency domain enforces $\mathbf{c}_{3,x}[\tau_1, \tau_2]$ and thus $\mathbf{c}_{3,y}[\tau_1, \tau_2]$ to have unlimited support in the time domain, it is reasonable to assume that the supports of $\mathbf{c}_{3,x}[\tau_1, \tau_2]$ and $\mathbf{c}_{3,y}[\tau_1, \tau_2]$ are limited within a certain range $-L \leq \tau_1, \tau_2 \leq L$ because those $\mathbf{c}_{3,x}[\tau_1, \tau_2]$ and $\mathbf{c}_{3,y}[\tau_1, \tau_2]$ outside such a range can be negligible in many real-world applications.
By stacking values $\mathbf{c}_{3,y}[\tau_1, \tau_2]$ and $\mathbf{c}_{3,x}[\tau_1, \tau_2]$ for $-L \leq \tau_1, \tau_2 \leq L$ respectively, we obtain vectors $\mathbf{c}_{3,y}\in \mathbb{R}^{(2L+1)^2M^3\times 1}$ and $\mathbf{c}_{3,x}\in \mathbb{R}^{(2L+1)^2N^2\times 1}$ as follows:
\begin{equation}
\begin{aligned}
\mathbf{c}_{3,y}&=[\mathbf{c}^T_{3,y}[L, L], \mathbf{c}^T_{3,y}[L-1, L], \cdots, \mathbf{c}^T_{3,y}[-L, L],\\
& \mathbf{c}^T_{3,y}[L, L-1], \mathbf{c}^T_{3,y}[L-1, L-1], \cdots, \mathbf{c}^T_{3,y}[-L, L-1],\\ 
&\cdots, \cdots, \cdots, \\
&\mathbf{c}^T_{3,y}[L, -L], \mathbf{c}^T_{3,y}[L-1, -L], \cdots, \mathbf{c}^T_{3,y}[-L, -L]]^T,
\end{aligned} \label{long y}
\end{equation}
\begin{equation}
\begin{aligned}
\mathbf{c}_{3,x}&=[\mathbf{c}^T_{3,x}[L, L], \mathbf{c}^T_{3,x}[L-1, L], \cdots, \mathbf{c}^T_{3,x}[-L, L],\\
& \mathbf{c}^T_{3,x}[L, L-1], \mathbf{c}^T_{3,x}[L-1, L-1], \cdots, \mathbf{c}^T_{3,x}[-L, L-1],\\
&\cdots, \cdots, \cdots, \\
&\mathbf{c}^T_{3,x}[L, -L], \mathbf{c}^T_{3,x}[L-1, -L], \cdots, \mathbf{c}^T_{3,x}[-L, -L]]^T.
\end{aligned} \label{long x}
\end{equation}
The reconstruction of $c_{3,x}(\tau_1,\tau_2)$ can be achieved by taking advantage of all significant lags of cross-cumulants among the compressive outputs in (\ref{long y}) 
as long as the linear relationship between (\ref{long y}) and (\ref{long x}) is figured out. To that end, several key observations are in order. First, given the definition of  $\mathbf{c}_{3,x}[\tau_1, \tau_2]$ in (\ref{c3x}) and the fact that $\mathbf{c}_{3,x}[\tau_1, \tau_2]$ has finite support within $-L \leq \tau_1, \tau_2 \leq L$, the support of $c_{3,x}(\tau_1, \tau_2)$ should be limited within $-LN \leq \tau_1, \tau_2 \leq LN$. As a result, the last $N^2-1$ entries in the vector $\mathbf{c}_{3,x}[L, L]$ are zero; for $\forall \tau_1$, the elements in the vector $\mathbf{c}_{3,x}[\tau_1, L]$ are all zero except those indexed by $jN+1, j=0,...,N-1$; for $\forall \tau_2$, the elements in the vector $\mathbf{c}_{3,x}[L, \tau_2]$ are all zero except the first $N$ elements. Similarly, given the definition in (\ref{c3e}) along with the fact that $c_{e_{i_1}, e_{i_2}, e_{i_3}}(\tau_1, \tau_2)$ has limited support within $1-N \leq \tau_1, \tau_2 \leq N-1$, the columns of matrix $\mathbf{C}_{3,e}[0,1]$ indexed by $jN+1, j=0,...,N-1$ are all zero; the first $N$ columns of matrix $\mathbf{C}_{3,e}[1,0]$ are all zero; the first $N$ columns and other columns of matrix $\mathbf{C}_{3,e}[1,1]$ that indexed by $jN+1, j=1,...,N-1$ are all zero. By capitalizing on these crucial observations, we can rewrite the two-dimensional linear convolution in (\ref{matrix conv}) into a one-dimensional circular convolution, which further allows us to figure out the relation between $\mathbf{c}_{3,y}$ in (\ref{long y}) and $\mathbf{c}_{3,x}$ in (\ref{long x}) as
\begin{equation}
\mathbf{c}_{3,y}=\mathbf{C}_{3,e}\mathbf{c}_{3,x} \label{circular conv}
\end{equation}
where $\mathbf{C}_{3,e}\in \mathbb{R}^{(2L+1)^2M^3\times (2L+1)^2N^2}$ is given by (\ref{C3e}). 
\begin{table*}[ht]
\begin{equation}
\begin{aligned}
\mathbf{C}_{3,e}=\left[
  \begin{array}{cccccccccc} 
    \mathbf{C}_{3,e}[0,0] & \mathbf{C}_{3,e}[1,0] &  &  & \mathbf{C}_{3,e}[0,1] & \mathbf{C}_{3,e}[1,1] & &\\ 
   & \mathbf{C}_{3,e}[0,0] & \mathbf{C}_{3,e}[1,0] &  &  & \mathbf{C}_{3,e}[0,1] & \mathbf{C}_{3,e}[1,1] & \\ 
   & &\ddots & \ddots & & & \ddots & \ddots  & \\ 
   & && \mathbf{C}_{3,e}[0,0] & \mathbf{C}_{3,e}[1,0]  &  & & \mathbf{C}_{3,e}[0,1] & \mathbf{C}_{3,e}[1,1]\\ 
   \mathbf{C}_{3,e}[1,1] && && \mathbf{C}_{3,e}[0,0] & \mathbf{C}_{3,e}[1,0]  &  & & \mathbf{C}_{3,e}[0,1]\\ 
    \mathbf{C}_{3,e}[0,1]&\mathbf{C}_{3,e}[1,1] && && \mathbf{C}_{3,e}[0,0] & \mathbf{C}_{3,e}[1,0]  & \\ 
    &\ddots & \ddots & & & & \ddots & \ddots  \\ 
    & & \mathbf{C}_{3,e}[0,1]&\mathbf{C}_{3,e}[1,1] && && \mathbf{C}_{3,e}[0,0] &\mathbf{C}_{3,e}[1,0] \\ 
      \mathbf{C}_{3,e}[1,0]   && & \mathbf{C}_{3,e}[0,1]&\mathbf{C}_{3,e}[1,1] && && \mathbf{C}_{3,e}[0,0]\\ 
  \end{array} 
\right]
\end{aligned} \label{C3e}
\end{equation}
\end{table*} Evidently, recovering $\mathbf{c}_{3,x}$ boils down to solving the linear system (\ref{circular conv}), which is invertible if $\mathbf{C}_{3,e}$ has full column rank.


By noting that $\mathbf{C}_{3,e}$ in (\ref{C3e}) is a block circulant matrix with blocks of size $M^3\times N^2$, the computational load and memory requirements for solving (\ref{circular conv}) can be greatly reduced. Specifically, it is known that block circulant matrices can be diagonalized by discrete Fourier transform (DFT) matrices \cite{khoromskaia2017block}. By using the $(2L+1)^2$-point (inverse) discrete Fourier transform ((I)DFT), the block circulant matrix $\mathbf{C}_{3,e}$ can be represented as
\begin{equation}
\mathbf{C}_{3,e} = (\mathbf{F}_{(2L+1)^2}^{-1}\otimes \mathbf{I}_{M^3}) \mathbf{Q}_{3e} (\mathbf{F}_{(2L+1)^2}\otimes  \mathbf{I}_{N^2})
\end{equation}
where $\mathbf{F}_{(2L+1)^2}$ is the $(2L+1)^2 \times (2L+1)^2$ DFT matrix, and $\mathbf{Q}_{3e} := diag\{\mathbf{Q}_{3e}(0), \mathbf{Q}_{3e}(2\pi \frac{1}{(2L+1)^2}), \cdots, \mathbf{Q}_{3e}(2\pi \frac{(2L+1)^2-1}{(2L+1)^2})\}$ is a block diagonal matrix with each block being of size $M^3\times N^2$.
Such a fact allows us to rewrite (\ref{circular conv}) as
\begin{equation}
\mathbf{q}_y= \mathbf{Q}_{3e} \mathbf{q}_x, \label{qy Q qx}
\end{equation} where the vectors $\mathbf{q}_y\in \mathbb{R}^{(2L+1)^2M^3\times 1}$ and $\mathbf{q}_x \in \mathbb{R}^{(2L+1)^2N^2\times 1}$ are given by
\begin{equation}
\mathbf{q}_y=(\mathbf{F}_{(2L+1)^2} \otimes  \mathbf{I}_{M^3})\mathbf{c}_{3,y}, \label{time-frequency y}
\end{equation}
\begin{equation}
\mathbf{q}_x=(\mathbf{F}_{(2L+1)^2} \otimes  \mathbf{I}_{N^2})\mathbf{c}_{3,x}. \label{time-frequency x}
\end{equation}
Plugging (\ref{long y}) and (\ref{long x}) into (\ref{time-frequency y}) and (\ref{time-frequency x}) yields
\begin{equation}
\mathbf{q}_y=[\mathbf{q}^T_y(0), \mathbf{q}^T_y(2\pi \frac{1}{(2L+1)^2}), \cdots,  \mathbf{q}^T_y(2\pi \frac{(2L+1)^2-1}{(2L+1)^2})]^T \label{structure qy}
\end{equation}
\begin{equation}
\mathbf{q}_x=[\mathbf{q}^T_x(0), \mathbf{q}^T_x(2\pi \frac{1}{(2L+1)^2}), \cdots,  \mathbf{q}^T_x(2\pi \frac{(2L+1)^2-1}{(2L+1)^2})]^T \label{structure qx}
\end{equation}
with $\mathbf{q}_y(\cdot)\in \mathbb{R}^{M^3\times 1}$ and $\mathbf{q}_x(\cdot)\in \mathbb{R}^{N^2 \times 1}$ given by:
\begin{equation}
\mathbf{q}_y(2\pi \frac{\ell}{(2L+1)^2}) =\left[\mathbf{F}_{(2L+1)^2} \otimes  \mathbf{I}_{M^3}\right]_{{\ell M^3+1} : (\ell+1)M^3,:} \mathbf{c}_{3,y} \label{qy fre}
\end{equation}
\begin{equation}
\mathbf{q}_x(2\pi \frac{\ell}{(2L+1)^2}) =\left[\mathbf{F}_{(2L+1)^2} \otimes  \mathbf{I}_{N^2}\right]_{{\ell N^2+1} : (\ell+1)N^2,:} \mathbf{c}_{3,x} \label{qx fre}
\end{equation}
where $\ell=0,\cdots,(2L+1)^2-1$ and $\left[\mathbf{F}_{(2L+1)^2} \otimes  \mathbf{I}_{M^3}\right]_{{a:b} ,:}$ denotes a submatrix. The block structure in (\ref{structure qy}) and (\ref{structure qx}) enables us to decompose (\ref{qy Q qx}) into $(2L+1)^2$ matrix equations:
\begin{equation}
\mathbf{q}_y(2\pi \frac{\ell}{(2L+1)^2})= \mathbf{Q}_{3e}(2\pi \frac{\ell}{(2L+1)^2}) \mathbf{q}_x(2\pi \frac{\ell}{(2L+1)^2}). \label{qyl qxl}
\end{equation}
Note that given $\mathbf{q}_y(2\pi \frac{\ell}{(2L+1)^2})$, the linear systems (\ref{qyl qxl}) for $\ell=0,\cdots,(2L+1)^2-1$ can be solved via LS  in a parallel fashion, as long as $\mathbf{Q}_{3e}(2\pi \frac{\ell}{(2L+1)^2})$ has full column rank. Once $\mathbf{q}_x$ is obtained, $ \mathbf{c}_{3,x}$ can be recovered based on (\ref{time-frequency x}).

To summarize the implementation of the proposed direct C3CS approach, firstly, $\mathbf{c}_{3,y}$ is calculated based on (\ref{cy and cz}) and (\ref{long y}) and $\mathbf{C}_{3,e}$ is constructed based on (\ref{c3ecumu}) and (\ref{C3e}); secondly, $\mathbf{q}_y$ is obtained according to (\ref{time-frequency y}) and (\ref{structure qy}); and finally $\mathbf{q}_x$ can be recovered by solving the linear system (\ref{qy Q qx}) or (\ref{qyl qxl}), which returns $ \mathbf{c}_{3,x}$ based on the relation (\ref{time-frequency x}).

\subsection{The Alternative C3CS Approach}
\label{ssec:subhead}

Different from the direct C3CS approach derived above, an alternative C3CS approach is presented in this subsection. First, by defining the $M\times 1$ measurement vectors $\mathbf{y}[k]$ and the $N\times 1$ vector sequence $\mathbf{x}[k]$ as
\begin{equation}
\mathbf{y}[k]=[y_0[k], y_1[k], ..., y_{M-1}[k]]^T, \label{y_alter}
\end{equation}
\begin{equation}
\mathbf{x}[k]=[x[kN], x[kN+1], ..., x[kN+N-1]]^T, \label{x_alter}
\end{equation}
we  rewrite (\ref{sampling device rewritten}) in the matrix-vector form
\begin{equation}
\mathbf{y}[k]=\mathbf{\Phi}\mathbf{x}[k], \label{yphix}
\end{equation}
where compressive sampling matrix $\mathbf{\Phi}\in \mathbb{R}^{M\times N}$ is given by
\begin{equation}
\mathbf{\Phi}=[\mathbf{e}_0, \mathbf{e}_1, \mathbf{e}_2, ..., \mathbf{e}_{M-1}]^T \label{phie}
\end{equation}
with $\mathbf{e}_i=[e_{i}[0], e_{i}[-1], ..., e_{i}[1-N]]^T$.

Next, we compute the three-way cumulant tensor $\mathbfcal{C}_{3,y}\in 
\mathbb{R}^{M\times M \times M}$ of $\mathbf{y}[k]$ in (\ref{y_alter}) by using tensor outer product $\mathbfcal{C}_{3,y}=E\left\lbrace\mathbf{y}[k]\circ \mathbf{y}[k]\circ \mathbf{y}[k]\right\rbrace$. It is obvious that $\mathbfcal{C}_{3,y}$ contains all third-order cumulants that can be calculated from $\mathbf{y}[k]$. Similarly, we can also construct the cumulant tensor $\mathbfcal{C}_{3,x} \in \mathbb{R}^{N\times N\times N}$ of $\mathbf{x}[k]$ in (\ref{x_alter}) as $\mathbfcal{C}_{3,x}=E\left\lbrace\mathbf{x}[k]\circ \mathbf{x}[k]\circ \mathbf{x}[k]\right\rbrace$. By using multilinear algebra, the relationship between $\mathbfcal{C}_{3,y}$ and $\mathbfcal{C}_{3,x}$ can be expressed as 
\begin{equation}
\mathbfcal{C}_{3,y}=\mathbfcal{C}_{3,x}\bullet_1\mathbf{\Phi}\bullet_2\mathbf{\Phi}\bullet_3\mathbf{\Phi} \label{tensor rep}
\end{equation}
where $\bullet_i$ represents the $i$th mode product between a tensor and a matrix \cite{cichocki2015tensor}. By vectorizing $\mathbfcal{C}_{3,y}$, it is evident that (\ref{tensor rep}) can be rewritten as
\begin{equation}
\begin{aligned}
\text{vec}(\mathbfcal{C}_{3,y}) &=E\left\lbrace(\mathbf{\Phi}\mathbf{x}[k])\circ (\mathbf{\Phi}\mathbf{x}[k])\circ (\mathbf{\Phi}\mathbf{x}[k])\right\rbrace\\
&= (\mathbf{\Phi} \otimes \mathbf{\Phi} \otimes \mathbf{\Phi}) \text{vec}(\mathbfcal{C}_{3,x}). \label{vectorize tensor}
\end{aligned}
\end{equation}
Thus far, the linear relationship between cross-cumulants of compressive samples and the cumulants of the uncompressed signal is explicitly figured out.

Alternatively, the third-order cumulants of $\mathbf{y}[k]$ and $\mathbf{x}[k]$ can also be calculated directly in the vector form as $\bar{\mathbf{c}}_{3,\mathbf{y}}=E\left\lbrace\mathbf{y}^{(3)}\right\rbrace$ of dimension $\mathbb{R}^{M^3 \times 1}$ and $\bar{\mathbf{c}}_{3,\mathbf{x}}=E\left\lbrace\mathbf{x}^{(3)}\right\rbrace$ of dimension $\mathbb{R}^{N^3 \times 1}$. By using the linear transformation property of Kronecker product $(\mathbf{\Phi}\mathbf{x})\otimes(\mathbf{\Phi}\mathbf{x})\otimes(\mathbf{\Phi}\mathbf{x})=(\mathbf{\Phi}\otimes\mathbf{\Phi}\otimes\mathbf{\Phi})(\mathbf{x}\otimes\mathbf{x}\otimes\mathbf{x})$ or equivalently $(\mathbf{\Phi}\mathbf{x})^{(3)}=\mathbf{\Phi}^{(3)}\mathbf{x}^{(3)}$ \cite{andre1997low}, $\bar{\mathbf{c}}_{3,\mathbf{y}}$ and $\bar{\mathbf{c}}_{3,\mathbf{x}}$ are related by
\begin{equation}
\begin{aligned}
\bar{\mathbf{c}}_{3,\mathbf{y}}&=E\left\lbrace\left(\mathbf{\Phi x}\right)^{(3)}\right\rbrace=\mathbf{\Phi}^{(3)}E\left\lbrace\mathbf{x}^{(3)}\right\rbrace=\mathbf{\Phi}^{(3)}\bar{\mathbf{c}}_{3,\mathbf{x}}. \label{cumulant relation 1}
\end{aligned}
\end{equation}
Actually, $\bar{\mathbf{c}}_{3,\mathbf{y}} =\text{vec}(\mathbfcal{C}_{3,y}) =  \mathbf{c}_{3,y}[0,0]$ and $\bar{\mathbf{c}}_{3,\mathbf{x}}=\text{vec}(\mathbfcal{C}_{3,x}) = \mathbf{c}_{3,x}[0,0]$, where $\mathbf{c}_{3,y}[0,0]$ and $\mathbf{c}_{3,x}[0,0]$ are defined in (\ref{long y}) and (\ref{long x}) respectively. Therefore, the linear systems (\ref{vectorize tensor}) and (\ref{cumulant relation 1}) are equivalent.

Since $\mathbf{\Phi}^{(3)}$ of size $M^3\times N^3$ is rank-deficient, the linear system in (\ref{cumulant relation 1}) is under-determined, and hence one cannot recover $\bar{\mathbf{c}}_{3,\mathbf{x}}$ accurately from $\bar{\mathbf{c}}_{3,\mathbf{y}}$ if not given any further constraints. Fortunately, it turns out that $\bar{\mathbf{c}}_{3,\mathbf{x}}$, or equivalently $\text{vec}(\mathbfcal{C}_{3,x})$, can actually be represented in a subspace with much smaller dimension, which consequently transforms the seemingly under-determined system (\ref{cumulant relation 1}) to be an over-determined one. To be specific, 
by noticing that the entry $\mathbfcal{C}_{3,x}(i,j,l)=E(x[kN+i]x[kN+j]x[kN+l]$ amounts to $c_{3,x}(j-i,l-i)$ according to (\ref{3 stationary cumulants}), we can form an index set $\mathbb{S}_{\mathbf{c}_{3,x}}$ with elements $(j-i,l-i)$ for $i,j,l=0,\cdots,N-1$ by removing repeated indices, and the cardinalty $|\mathbb{S}_{\mathbf{c}_{3,x}}|$ equals to the degrees of freedom contained in $\mathbfcal{C}_{3,x}$. It can be easily verified that $|\mathbb{S}_{\mathbf{c}_{3,x}}|=3N^2-3N+1$. Since tensor $\mathbfcal{C}_{3,x}$ with $N^3$ elements contains only $3N^2-3N+1$ degrees of freedom, $\mathbfcal{C}_{3,x}$ can be condensed into a $(3N^2-3N+1) \times 1$ vector $\mathbf{c}_x^a$ whose elements are $c_{3,x}(\tau_1, \tau_2), \forall (\tau_1, \tau_2) \in \mathbb{S}_{\mathbf{c}_{3,x}}$ and we can write
\begin{equation}
\bar{\mathbf{c}}_{3,\mathbf{x}} = \text{vec}(\mathbfcal{C}_{3,x}) = \mathbf{T}_N\mathbf{c}_x^a \label{subspace tensor}
\end{equation}
where $\mathbf{T}_N$ is a special $N^3\times (3N^2-3N+1)$ mapping matrix. We will show how such deterministic mapping matrix can be constructed in the next section after accounting for further information.

Plugging (\ref{subspace tensor}) into (\ref{cumulant relation 1}) yields
\begin{equation}
\bar{\mathbf{c}}_{3,\mathbf{y}} =  (\mathbf{\Phi} \otimes \mathbf{\Phi} \otimes \mathbf{\Phi}) \mathbf{T}_N\mathbf{c}_x^a=\mathbf{\Phi}_c^a \mathbf{c}_x^a \label{least square formu tensor}
\end{equation}
where $\mathbf{\Phi}_c^a = (\mathbf{\Phi} \otimes \mathbf{\Phi} \otimes \mathbf{\Phi})\mathbf{T}_N$ is of dimension $M^3\times (3N^2-3N+1)$. Evidently, obtaining $\mathbf{c}_x^a$ is equivalent to recovering $\bar{\mathbf{c}}_{3,\mathbf{x}}$ because both of them contain the same set of values. The unique recovery of $\mathbf{c}_x^a$ is guaranteed by the following theorem.

\textbf{\emph{Theorem 1 (Sufficient Conditions)}}: $\mathbf{c}_x^a$ can be losslessly recovered from $\bar{\mathbf{c}}_{3,\mathbf{y}}$ in the linear system (\ref{least square formu tensor}) via solving simple LS, if one of the following two conditions holds: 

\emph{1)} The matrix $\mathbf{\Phi}_c^a$ is deterministic and full column rank.

\emph{2)}  The matrix $\mathbf{\Phi} \in \mathbb{R}^{M\times N}$ is random (Guassian random matrix for example) and $(M+2)(M+1)M \geq 6(3N^2-3N+1)$, such that \emph{1)} is satisfied with high probability. 

\begin{proof}: It is straightforward to see that condition \emph{1)} is sufficient for lossless recovery of $\mathbf{c}_x^a$ . We directly go to examine condition \emph{2)}, which is more practical for the compressive sampling matrix design. To figure out if and when the linear system (\ref{least square formu tensor}) is over-determined, we need to determined the degrees of freedom contained in $\bar{\mathbf{c}}_{3,\mathbf{y}}$. Observing the compressive sampling strategy (\ref{yphix}), each element of $\mathbf{y}[k]$ is generated by filtering $\mathbf{x}[k]$ with an individual filter $\phi_k$ indicating by the corresponding row of $\mathbf{\Phi}$, where $\{\phi_k\}_k$ are mutually incoherent with high probability if a random Guassian matrix is used. As a result, $\mathbf{y}[k]$ is not stationary, although $\mathbf{x}[k]$ is, which implies that the third-order cumulant of $\mathbf{y}[k]$ is not only dependent on time lags $\tau_1$ and $\tau_2$, but also dependent on time origin $t$. Therefore, the redundant values in $\bar{\mathbf{c}}_{3,\mathbf{y}}$ come only from the inherent symmetry of the Kronecker product, e.g., $c_{3,y}(t; \tau_1, \tau_2)=cum(y(t), y(t+\tau_1), y(t+\tau_2))=cum(y(t), y(t+\tau_2), y(t+\tau_1))=c_{3,y}(t; \tau_2, \tau_1)$ \cite{meijer2005matrix}. In that case, the number of unique elements contained in $\bar{\mathbf{c}}_{3,\mathbf{y}}$ is actually $\tbinom{3+M-1}{3}=\frac{(3+M-1)!}{3!(M-1)!}=\frac{(M+2)(M+1)M}{6}$. In addition, the degrees of freedom in $\mathbf{c}_x^a$ is $3N^2-3N+1$ as analyzed in (\ref{subspace tensor}). Consequently, we conclude that as long as $\frac{(M+2)(M+1)M}{6}\geq3N^2-3N+1$, the linear system (\ref{least square formu tensor}) is over-determined with high probability and a unique solution is permitted via solving simple LS. 
\end{proof}

%
%
%

Theorem 1 provides sufficient conditions on matrices $\mathbf{\Phi}_c^a$ and $\mathbf{\Phi}$ to guarantee lossless recovery, and it should be noted that the elements of $\mathbf{\Phi}$ are determined by $\mathbf{e}_i$ as shown in (\ref{phie}). That is to say, $e_i(t)$, or equivalently $p_i(t)$, should be carefully chosen according to Theorem 1 so that the mutual incoherence of matrix $\mathbf{\Phi}$ can be guaranteed with high probability. Besides, the above sufficient condition \emph{2)} fundamentally reveals the strongest achievable compression when using the proposed alternative C3CS approach. Specifically, the ratio $M/N$ reflects the saving of the sampling cost and the strongest compression is given when the equality in condition \emph{2)} holds, that is, $(M+2)(M+1)M/6=3N^2-3N+1$ for a given value of $N$. A more straightforward illustration of compression efficiency will be provided in the next section. 
 In general, we choose $(M+1)(M+2)M/6$ to be slightly greater than $3N^2-3N+1$ to satisfy condition \emph{2)}, which results in the computational complexity of our method to be comparable to calculating cumulants directly from uncompressive measurements. 

\textbf{\emph{Remark 1 (comparison to the direct C3CS approach)}}: Generally speaking, the direct C3CS approach is able to estimate more cumulant lags than the alternative C3CS approach. The reason is that the direct C3CS approach takes advantage of all inter-data-blocks cross-cumulants $c_{y_{i_1}, y_{i_2}, y_{i_3}}( \tau_1, \tau_2)$, $\forall \tau_1,\tau_2= 0, 1, \cdots, L-1$ of compressive outputs, whereas the alternative C3CS approach considers only cross-cumulants within one data block, i.e., $c_{y_{i_1}, y_{i_2}, y_{i_3}}(0, 0)$. Once $L$ is chosen, all the cumulant values $c_{3x}(\tau_1, \tau_2), -LN \leq \tau_1, \tau_2 \leq LN$ can be estimated. Obviously, the larger $L$ is chosen, the more cumulant values can be estimated, which, however, also increases computational complexity because more cumulant values need to be calculated.


\textbf{\emph{Remark 2 (relation to CCS)}}: The multilinear system (\ref{tensor rep}) is closely related to CCS where cumulant tensors $\mathbfcal{C}_{3,y}$ and $\mathbfcal{C}_{3,x}$ reduce to covariance matrices $\mathbf{C}_{2,y}$ and $\mathbf{C}_{2,x}$ and they are related by $\mathbf{C}_{2,y} = \mathbf{C}_{2,x}\bullet_1\mathbf{\Phi}\bullet_2\mathbf{\Phi}=\mathbf{\Phi}\mathbf{C}_{2,x}\mathbf{\Phi}^T$. In CCS, it recovers $\mathbf{C}_{2,x}$ by capitalizing on its Toeplitz structure. The similarity between CCS and our approach is that it is the stationarity that results in the special structures of $\mathbf{C}_{2,x}$ and $\mathbfcal{C}_{3,x}$, and hence allows the reduction in the degrees of freedom to be estimated to achieve over-determined systems.

\textbf{\emph{Remark 3 (generalization to higher-order statistics)}}: Our proposed approach can be easily generalized to higher than third-order statistics. Specifically, the $q$th-order moments of $\mathbf{y}[k]$ and $\mathbf{x}[k]$ can be arranged into $q$-way tensors as $\mathbfcal{C}_{q,y} \in \mathbb{R}^{M\times M\times \cdots \times M}$ and $\mathbfcal{C}_{q,x} \in \mathbb{R}^{N\times N\times \cdots \times N}$, and they are related through a multilinear system like (\ref{tensor rep}). By capitalizing on the stationarity of $\mathbf{x}[k]$, it turns out that the degrees of freedom contained in $\mathbfcal{C}_{q,x}$ can be condensed into a vector of much smaller dimension like in (\ref{subspace tensor}). Then an over-determined system can be obtained similar to (\ref{least square formu tensor}). 

\section{Additional Symmetry Information}
\label{sec:majhead}

This section explores the use of symmetry information of third-order cumulants in order to improve both compression efficiency and estimation performance.

\textbf{\emph{Symmetry Property \cite{mendel1991tutorial}:}} The definition of the third-order cumulant of stationary random process in (\ref{3 stationary cumulants}) implies an important symmetry property: $c_{3,x}(\tau_1, \tau_2)=c_{3,x}(\tau_2, \tau_1)=c_{3,x}(-\tau_2,\tau_1- \tau_2)=c_{3,x}(-\tau_1, \tau_2-\tau_1)=c_{3,x}(\tau_2-\tau_1, -\tau_1)=c_{3,x}(\tau_1-\tau_2, -\tau_2)$.

 \begin{figure}
  \centering
    \centerline{\includegraphics[width=4cm, height=3.5cm]{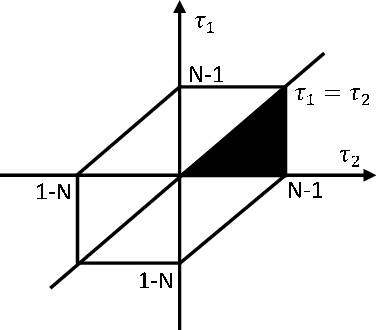}}
  \caption{\label{fig: cumulant support}The support regions of the third-order cumulant in the $\tau_1-\tau_2$ space.} 
\end{figure}

By using these symmetry equations, the $\tau_1\textrm{-} \tau_2$ plane can be divided into six triangular regions. Giving the cumulants in any region, the cumulants in the other five regions can be determined. The principal region is as shown in the shadowed sector in Fig. \ref{fig: cumulant support}, which is defined by $0\leq\tau_1\leq\tau_2\leq N$. The degrees of freedom to be estimated are completely determined by the elements located in the principal region.
Hence, for stationary random process $\mathbf{x}[k]$, the number of unique values contained in $\bar{\mathbf{c}}_{3,\mathbf{x}}$ is exactly $N(N+1)/2$. We stack these unique values in a vector denoted $\tilde{\mathbf{c}}_{3,\mathbf{x}} \in \mathbb{R}^{N(N+1)/2 \times 1}$ as follows
\begin{equation}
\begin{aligned}
\tilde{\mathbf{c}}_{3,\mathbf{x}} = &[c_{3,x}(0, 0), c_{3,x}(0, 1), \cdots, c_{3,x}(0, N-1),\\
&c_{3,x}(1, 1), c_{3,x}(1, 2), \cdots, c_{3,x}(1, N-1),\\
&\cdots \cdots, c_{3,x}(N-1, N-1)]^T.
\end{aligned} 
\end{equation}
Evidently, obtaining the values of $\tilde{\mathbf{c}}_{3,\mathbf{x}}$ is equivalent to recovering $\bar{\mathbf{c}}_{3,\mathbf{x}}$, as long as the relationship between them can be figured out explicitly.

 


\emph{Relating $\tilde{\mathbf{c}}_{3,\mathbf{x}}$ to $\bar{\mathbf{c}}_{3,\mathbf{x}}$:} Both $\tilde{\mathbf{c}}_{3,\mathbf{x}}$ to $\bar{\mathbf{c}}_{3,\mathbf{x}}$ contain the same set of nonzero elements, and hence can be linearly related to each other as:
 \begin{equation}
  \bar{\mathbf{c}}_{3,\mathbf{x}} = \mathbf{P}_N\tilde{\mathbf{c}}_{3,\mathbf{x}}, \label{unique cumulants}
 \end{equation}
where $\mathbf{P}_N  \in \left\lbrace0, 1\right\rbrace^{N^3 \times N(N+1)/2}$ is a deterministic mapping matrix constructed as follows:
\begin{equation}
\begin{cases}
\left[\mathbf{P}_N\right]_{\left(p_i(u,v,w), q(u,v)\right)} = 1, \quad  i = 1,2,...,6,\\
\left[\mathbf{P}_N\right]_{otherwise} = 0,
\end{cases}\\ \label{PN}
\end{equation}
where $\left[\mathbf{P}_N\right]_{(p,q)}$ denotes the $(p, q)$ element in matrix $\mathbf{P}_N$, and $p_i(u,v,w)$ and $q(u,v)$ are given by
\begin{equation}
\begin{aligned}
&p_1(u,v,w) = (w-1)N^2+(w+v-1)N+(w+u)\\
&p_2(u,v,w) = (w-1)N^2+(w+u-1)N+(w+v)\\
&p_3(u,v,w) = (w+v-1)N^2+(w-1)N+(w+u)\\
&p_4(u,v,w) = (w+v-1)N^2+(w+u-1)N+w\\
&p_5(u,v,w) = (w+u-1)N^2+(w-1)N+(w+v)\\
&p_6(u,v,w) = (w+u-1)N^2+(w+v-1)N+w\\
&q(u,v) = \begin{cases} 
      v-u+1 & u=0\\
      \sum_{j=1}^u(N-j+1)+v-u+1 & u\neq 0
   \end{cases}\\
&\forall u\in [0, N-1], v\in [u, N-1], w \in [1, N-v],
\end{aligned} \label{PN_detail}
\end{equation}
The derivation of $\mathbf{P}_N$ is given in Appendix A.

 
 \begin{figure}
  \centering
  \centerline{\includegraphics[width=9cm, height=5.5cm]{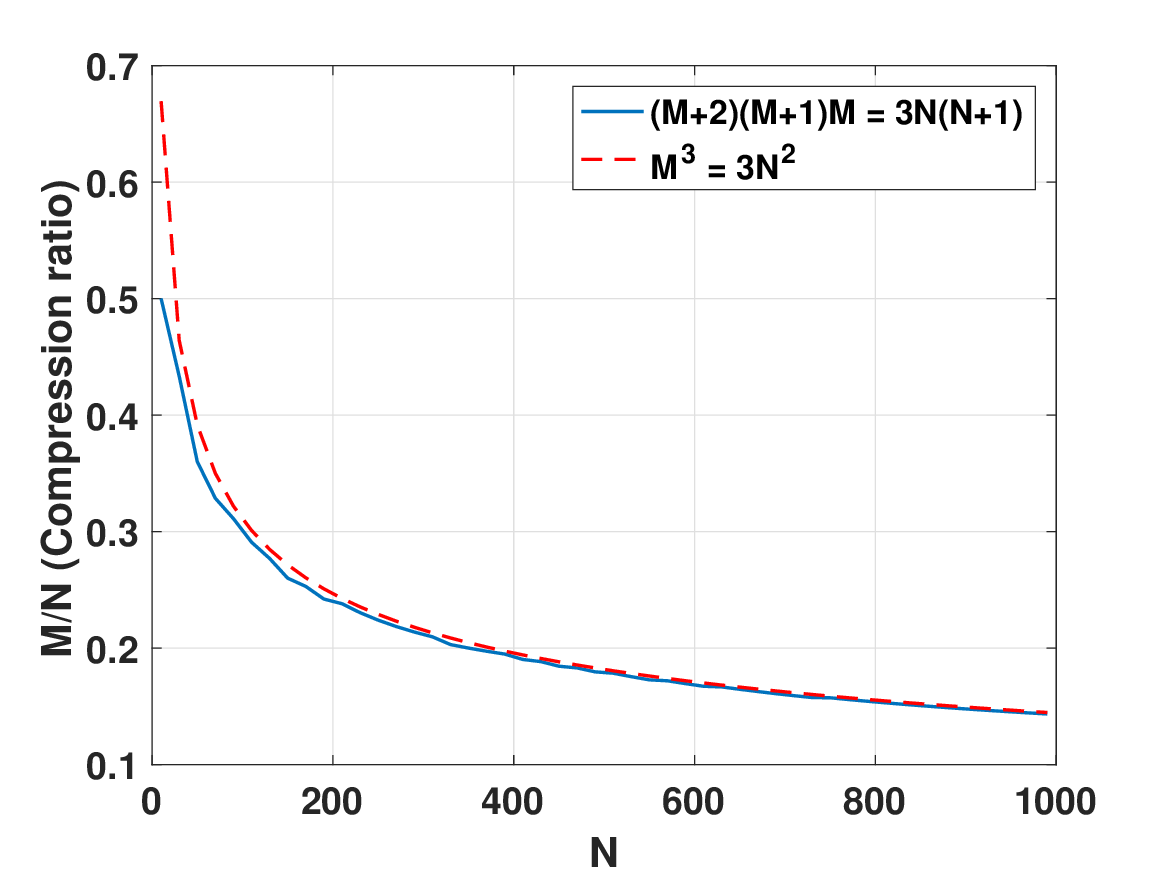}}
  \caption{\label{fig:  strongest compression}The strongest achievable compression ratio $[M/N]$ vs block length $N$.} 
\end{figure}

Plugging (\ref{unique cumulants}) into (\ref{cumulant relation 1}) yields:
\begin{equation}
\bar{\mathbf{c}}_{3,\mathbf{y}}=\mathbf{\Phi}^{(3)}\mathbf{P}_N\tilde{\mathbf{c}}_{3,\mathbf{x}}. \label{time domain relation}
\end{equation} where $ (\mathbf{\Phi} \otimes \mathbf{\Phi} \otimes \mathbf{\Phi})\mathbf{P}_N$ is of dimension $M^3\times N(N+1)/2$. 
It is worth noting that $\tilde{\mathbf{c}}_{3,\mathbf{x}}$ contains less degree of freedom to be estimated than $\mathbf{c}_x^a$, and hence stronger compression and better reconstruction performance can be expected. The unique recovery of $\tilde{\mathbf{c}}_{3,\mathbf{x}}$ is guaranteed by the following theorem.

\textbf{\emph{Theorem 2 (Sufficient Conditions)}}: $\tilde{\mathbf{c}}_{3,\mathbf{x}}$ can be losslessly recovered from $\bar{\mathbf{c}}_{3,\mathbf{y}}$ in the linear system (\ref{time domain relation}) via solving simple LS, if one of the following two conditions holds: 

\emph{1)} The matrix $(\mathbf{\Phi} \otimes \mathbf{\Phi} \otimes \mathbf{\Phi})\mathbf{P}_N\in \mathbb{R}^{M^3\times N(N+1)/2}$ is deterministic and full column rank.

\emph{2)} The matrix $\mathbf{\Phi} \in \mathbb{R}^{M\times N}$ is random (Guassian random matrix for example) and $(M+2)(M+1)M \geq 3N(N+1)$, such that \emph{1)} is satisfied with high probability. 

%
%
%

The proof for Theorem 2 is similar to that for Theorem 1. Theoretically, the strongest compression is achieved when the equality holds in \emph{2)}, say $(M+2)(M+1)M = 3N(N+1)$.  Evidently, Theorem 2 provides stronger compression than Theorem 1 due to the proper use of symmetry information. To gain intuition on the compression ratio of our proposed approach, Fig. \ref{fig:  strongest compression} shows how strongest achievable compression ratio $[M/N]$ varies as $N$ increases. We can see that the relation $(M+2)(M+1)M = 3N(N+1)$ can be approximated by $M^3=3N^2$ as $N$ increases, which suggests the strongest compression ratio
\begin{equation}
\left(\frac{M}{N}\right)_{min} \approx \frac{\sqrt[3]{3N^2}}{N} = \sqrt[3]{\frac{3}{N}}. \label{compression ratio}
\end{equation}
The approximation (\ref{compression ratio}) makes it evident that our proposed cumulant reconstruction approach can achieve significant compression even \emph{without any sparsity assumption}. As is shown in Fig. \ref{fig:  strongest compression}, the strongest compression ratio of our proposed cumulant reconstruction approach can be even lower than $0.2$ when $N$ is greater than $400$. We provide the achievable strongest compression ratio corresponding to   a variety of values of $N$ and $M$ in Table \ref{table:M/N}.

\begin{table}[ht]
\caption{The Strongest Compression Ratio versus $N$ and $M$}
\centering
\begin{tabular}{c c c}
\hline \hline
N & The smallest M & Compression ratio M/N \\ [0.5ex]
\hline
20 & 11 & 0.550 \\
40 & 17 & 0.425 \\
80 & 27 & 0.338 \\
160 & 43 & 0.269 \\
320 & 68 & 0.212 \\
\hline
\label{table:M/N}
\end{tabular}

\end{table}
%
%

\textbf{\emph{Remark 4 (compression efficiency comparison)}}: The achievable strongest compression (\ref{compression ratio}) for the third-order cumulant reconstruction in C3CS is weaker than that for covariance reconstruction in CCS which is $\sqrt{\frac{2}{N}}$ as given in \cite{tian2011cyclic}. This is intuitively understandable because the third-order statistics contain richer information than second-order statistics and thus weaker compression is permitted in order to preserve such useful information.

To summarize, the alternative C3CS approach accounting for cumulant symmetry information is composed of the compressive sampling matrix design according to Theorem 2 and a cumulant recovery method. Given $\bar{\mathbf{c}}_{3,\mathbf{y}}$, the mapping matrix $\mathbf{P}_N$ and the predefined sampling matrix $\mathbf{\Phi}$, the cumulant recovery problem can be formulated as a simple LS problem without any constraints for reconstructing $\bar{\mathbf{c}}_{3,x}$ as follows:
\begin{equation}
\mathop {\min }\limits_{\tilde{\mathbf{c}}_{3,x}} \norm{\bar{\mathbf{c}}_{3,\mathbf{y}}-\mathbf{\Phi}^{(3)}\mathbf{P}_N\tilde{\mathbf{c}}_{3,\mathbf{x}}}_2^2. \label{time domain formulation}
\end{equation}
The over-determined least-squares formulation in (\ref{time domain formulation}) leads to a closed-form solution. Although sparsity is not entailed in C3CS, the sparsity prior, if exists, can also be properly incorporated to enhance the recovery performance. In addition, a key observation is that the matrix $\mathbf{P}_N$ is extremely sparse with each row having only one non-zero element, which facilitates the solving of (\ref{time domain formulation}) in large scale settings.

\section{Diagonal Cumulant Slice Reconstruction}
\label{sec:format}

In many signal processing applications, the 1-D slice of the higher-order cumulant, defined as $c_{q, x}(\tau,\cdots,\tau)=cum\left\lbrace x(n)x(n+\tau)\cdots x(n+\tau) \right\rbrace$, is sufficient for inference tasks therein, such as harmonics retrieval \cite{swami1991cumulant}, channel estimation \cite{liang2004fir}, system identification \cite{giannakis1989identification}, to name just a few. In that case, it would be desirable to recover the 1-D cumulant slice only instead of all the cumulant values for the purpose of reducing computational complexity. To that end, 
this section proposes an approach termed compressive cumulant slice sensing (CCSS) to reconstruct the 1-D diagonal cumulant slice directly from compressive measurements. In CCSS, a sparse sampling strategy is adopted, which links the compressive outputs and the 1-D diagonal slice of cumulants explicitly demonstrating the possibility of lossless recovery of the 1-D diagonal slice straightforwardly. In general, CCSS is able to achieve significant compression for \emph{arbitrary}-order cumulant diagonal slice recovery. 

To motivate the adopted sparse sampling strategy in CCSS, we first introduce the structure embedded in the 1-D diagonal cumulant slice of the $q$th-order cumulant.
Recall the sampling system in (\ref{yphix}).
 An intuitive way of calculating the 1-D diagonal cumulant slice of $\mathbf{x}[k]$ is given by $c_{q, x}(\tau,\cdots,\tau)=cum(x[kN+i], x[kN+j],\cdots,x[kN+j]))$ for any $j-i = \tau$, 
 where $0 \leq i,j \leq N-1$ denote the indices of the elements in the vector $\mathbf{x}[k]$.
 Because of the stationarity of $\mathbf{x}[k]$, it always holds that $c_{q, x}(j-i,\cdots,j-i)=c_{q, x}(n-m,\cdots,n-m)$ as long as $n-m=j-i=\tau$, for $m\neq i$ and $n\neq j$ where $0 \leq m,n \leq N-1$, similar to $i, j$, also denote the indices of the elements in the vector $\mathbf{x}[k]$. By stacking the non-zero diagonal cumulant slice values of $\mathbf{x}[k]$ into a vector as $\underline{\mathbf{c}}_{q,x}=[\cdots, c_{q,x}(\tau, \cdots, \tau), \cdots]^T$, $\tau=1-N,...0,...,N-1$, the above analysis on the stationary property implies that not every value $x[kN+n], n=0,...,N-1$ in $\mathbf{x}[k]$ is required to calculate $\underline{\mathbf{c}}_{q, x}$, but merely those values indexed by the elements of a set $\mathbb{M}$ defined by $\mathbb{M}:=\left\lbrace m_0, m_1, ..., m_{M-1}\right\rbrace \subseteq \left\lbrace 0,...,N-1\right\rbrace$ such that $ \{ m_j - m_i : \forall m_i,m_j \in \mathbb{M}) \} =  \left\lbrace 1-N, ..., 0,...,N-1\right\rbrace$. 
This suggests that $\underline{\mathbf{c}}_{q, x}$ can be accurately estimated from a subset of entries of $\mathbf{x}[k]$, which enables appealing signal compression for coping with wideband signals. Evidently, the choice of $\mathbb{M}$, if exists, is crucial for compressive efficiency. It turns out that several solutions to $\mathbb{M}$ are available, including the minimal sparse ruler, \cite{romero2016compressive}, nested sampling and co-prime sampling \cite{pal2011coprime} \cite{pal2010nested}. This observation sheds light on the sub-Nyquist sampler design and the reconstruction of $\underline{\mathbf{c}}_{q, x}$ from compressive measurements.


\subsection{Sampling Strategy for CCSS}
\label{ssec:subhead}

We adopt the nonuniform sub-sampling to reduce the average sampling rate \cite{romero2016compressive}. Specifically, the sampling matrix $\boldsymbol{\Phi} \in  \mathbb{R}^{M\times N}$ in (\ref{yphix})  is designed as a sparse matrix with its elements indexed by $(i, m_i)$ ($i=0,...,M-1, m_i \in \mathbb{M}$) being ones and otherwise zeros, where $\mathbb{M}:=\left\lbrace m_0, m_1, ..., m_{M-1}\right\rbrace \subseteq \left\lbrace 0,...,N-1\right\rbrace$. As a result, only a small number of samples are collected into the $M\times1$ measurements vector $\mathbf{y}[k]$ in (\ref{y_alter}), the elements of which are indexed by a subset of the Nyquist grid:
\begin{equation}
y_{i}[k]= x[(kN+m_i)T_s], \qquad m_i \in \mathbb{M}, \label{compressive sampling procedure}
\end{equation}
where $i=0,\cdots,M-1$.

It should be noted that $\mathbf{x}[k]$ is wide-sense stationary but $\mathbf{y}[k]$ is generally not because the nature of the compressive sampling matrix $\boldsymbol{\Phi}$, and hence the third-order cumulant of $\mathbf{y}$ is dependent on both the time origin and time-lags. This fact makes it possible that $\underline{\mathbf{c}}_{q, y}$ and $\underline{\mathbf{c}}_{q, x}$ contain the same degree of freedom even though the dimension of $\mathbf{y}[k]$ is smaller than that of $\mathbf{x}[k]$, where $\underline{\mathbf{c}}_{q, y}$ stacks all time-variant cumulant slice values of $\mathbf{y}[k]$. 

To relate $\underline{\mathbf{c}}_{q,x}$ and $\mathbf{y}[k]$, it can be easily found that
\begin{equation}
\begin{aligned}
c_{q,x}[k_j-k_i]=&cum(x[kN+m_i], x[kN+m_j], ..., x[kN+m_j])\\
=&cum(y_i[k], y_j[k], ..., y_j[k])).
\end{aligned} \label{cumulant y and z}
\end{equation}
This simple relationship in (\ref{cumulant y and z}) fundamentally reveals that $\underline{\mathbf{c}}_{q,x}$ can be completely recovered from $\mathbf{y}[k]$,  as long as for every $\tau=1-N,...,N-1$, there exists at least one pair $m_i$, $m_j$ in $\mathbb{M}$ satisfying $m_i-m_j=\tau$. 

To achieve the strongest compression ratio, the compressive  sampler design basically boils down to determining the minimal set $\mathbb{M}$ such that $\underline{\mathbf{c}}_{q,x}$ can be estimated accurately from compressively collected samples indexed by $\mathbb{M}$. Mathematically, this is a minimal sparse ruler design problem:
\begin{equation}
\mathop {\min }\limits_{\mathbb{M}}|\mathbb{M}| \quad s.t. \quad \mathbb{D}=\left\lbrace1-N,...,0,...,N-1\right\rbrace, \label{sparse ruler}
\end{equation}
where $\mathbb{D}=\left\lbrace m_j-m_i :  \forall m_i, m_j \in \mathbb{M} \right\rbrace$ and $|\mathbb{M}|$ denotes the cardinality of the set $\mathbb{M}$.
The minimal sparse ruler problem is well studied in \cite{leech1956representation} and has many precomputed solutions at hand for our design of the sparse sampling matrix $\boldsymbol{\Phi}$.  
 In practice, $\boldsymbol{\Phi}$ is composed of the rows of identity matrix $\mathbf{I}_N$ indexed by the elements of $\mathbb{M}$.
 
\subsection{Cumulant Slice Reconstruction}
\label{ssec:subhead}

Assume that multiple blocks of data records ($K$ blocks) are available and each record is denoted by $\mathbf{x}[k], k=1,\cdots,K$. Adopting the same sampler $\mathbf{\Phi} \in  \mathbb{R}^{M\times N}$ ($M<N$) for all blocks, the compressive sample vector $\mathbf{y}[k] \in  \mathbb{R}^{M\times 1}$ is obtained as
$\mathbf{y}[k]=\boldsymbol{\Phi}\mathbf{x}[k]$, $\forall k$. 
Based on (\ref{cumulant y and z}), the finite-sample estimate of the cumulant slice is obtained by averaging over all records:
\begin{equation}
\hat{c}_{q,x}(\tau, \cdots, \tau)=\frac{1}{K}\sum_{k=1}^B\hat{c}_{q,x_k}^{(k)}(\tau, \cdots, \tau), \label{average over data blocks}
\end{equation}
where $\hat{c}_{q,x_k}^{(k)}(\tau, \cdots, \tau)$ is the estimate based on the $k$th subrecord $\mathbf{y}[k]$. For example, the second-, third-and fourth-order ($q=2, 3, 4$) unbiased estimate are:
\begin{equation}
\begin{aligned}
\hat{c}_{2, x_k}^{(b)}(\tau)=& \frac{1}{\#}\sum_{i,j}x[kN+m_i]x[kN+m_j]\\
&=\frac{1}{\#}\sum_{i,j}y_i[k]y_j[k] \label{2nd_est} \\
\end{aligned}
\end{equation}
\begin{equation}
\begin{aligned}
\hat{c}_{3, x_k}^{(b)}(\tau, \tau)=& \frac{1}{\#}\sum_{i,j}x[kN+m_i]x[kN+m_j]^2 \\
=&\frac{1}{\#}\sum_{i,j}y_i[k]y_j[k]^2 \label{3rd_est} \\
\end{aligned}
\end{equation}
\begin{equation}
\begin{aligned}
\hat{c}_{4, x_k}^{(b)}(\tau, \tau, \tau)=&\frac{1}{\#}\sum_{i,j}x[kN+m_i]x[kN+m_j]^3\\
&-3\frac{1}{\#^2}\sum_{i,j}x[kN+m_i]x[kN+m_j]\\
&\qquad \qquad \qquad \sum_{i,j}x[kN+m_j]x[kN+m_j]\\
=&\frac{1}{\#}\sum_{i,j}\mathbf{y}_b[i]\mathbf{y}_b[j]\mathbf{y}_b[j]\mathbf{y}_b[j]\\
&-3\frac{1}{\#}\sum_{i,j}\mathbf{y}_b[i]\mathbf{y}_b[j]\frac{1}{\#}\sum_{i,j}\mathbf{y}_b[j]\mathbf{y}_b[j], \label{4th_est}
\end{aligned}
\end{equation}
where $\#$ denotes the number of averaged terms including all $i,j$ satisfying $m_j-m_i=\tau$
for all $\tau=1-N,...,0,...,N-1$. 

Evidently, CCSS is able to accurately recover arbitrary $q$th-order cumulant slice $\underline{\mathbf{c}}_{q, x}$ ($q\geq2$) from compressive measurements, which subsumes CCS as a special case with $q=2$. The compression ratio of sparse ruler based CCSS should be the same as that of CCS because both of them depends only on the cardinalty of $\mathbb{M}$, the analysis of which can be found in detail in \cite{romero2015compression}. The computation complexity of cumulant slice estimation is comparable to that of covariance estimation as is analyzed in \cite{swami1991cumulant}, which is much lighter than the entire cumulant estimation.

\section{SIMULATIONS}
\label{sec:majhead}
%

This section presents simulation results to corroborate the effectiveness of the proposed approaches on the sampling rate reduction as well as colored Gaussian noise suppression. Specific applications including correlation function estimation and line spectrum estimation are considered. 

\subsection{C3CS: Robustness to Sampling Rate Reduction}
\label{ssec:subhead}
The discrete-time data $x(n)$ used for simulation is generated by passing a zero-mean, independent and exponentially distributed driven process $w(n)$ through an MA linear system with coefficients [1, 0.9, 0.385,-0.771], say $x(n)=w(n)+0.9w(n-1)+0.385w(n-2)-0.771w(n-3)$, which has a finite number of cumulant and correlation values. We process $x(n)$ block by block with each data block $\mathbf{x}[k], k=1,...,K$ of equal length $N=20$. The compressive measurements $\mathbf{y}[k]$ of size $M\times 1$ is obtained from (\ref{yphix}), and the value of $M$ that decides the compression ratio will be specified in each simulation case. As analyzed in the Remark 1, given fixed $N$, the proposed direct C3CS approach is able to estimate cumulants within lags $-LN\leq \tau_1, \tau_2 \leq LN$, while the alternative C3CS approach within $1-N\leq \tau_1, \tau_2 \leq N-1$. In this simulation, since cumulants at lags outside $1-N\leq \tau_1, \tau_2 \leq N-1$ is negligible, we test the alternative C3CS approach directly.

\begin{figure}
  \centering
  \centerline{\includegraphics[width=9cm, height=5.5cm]{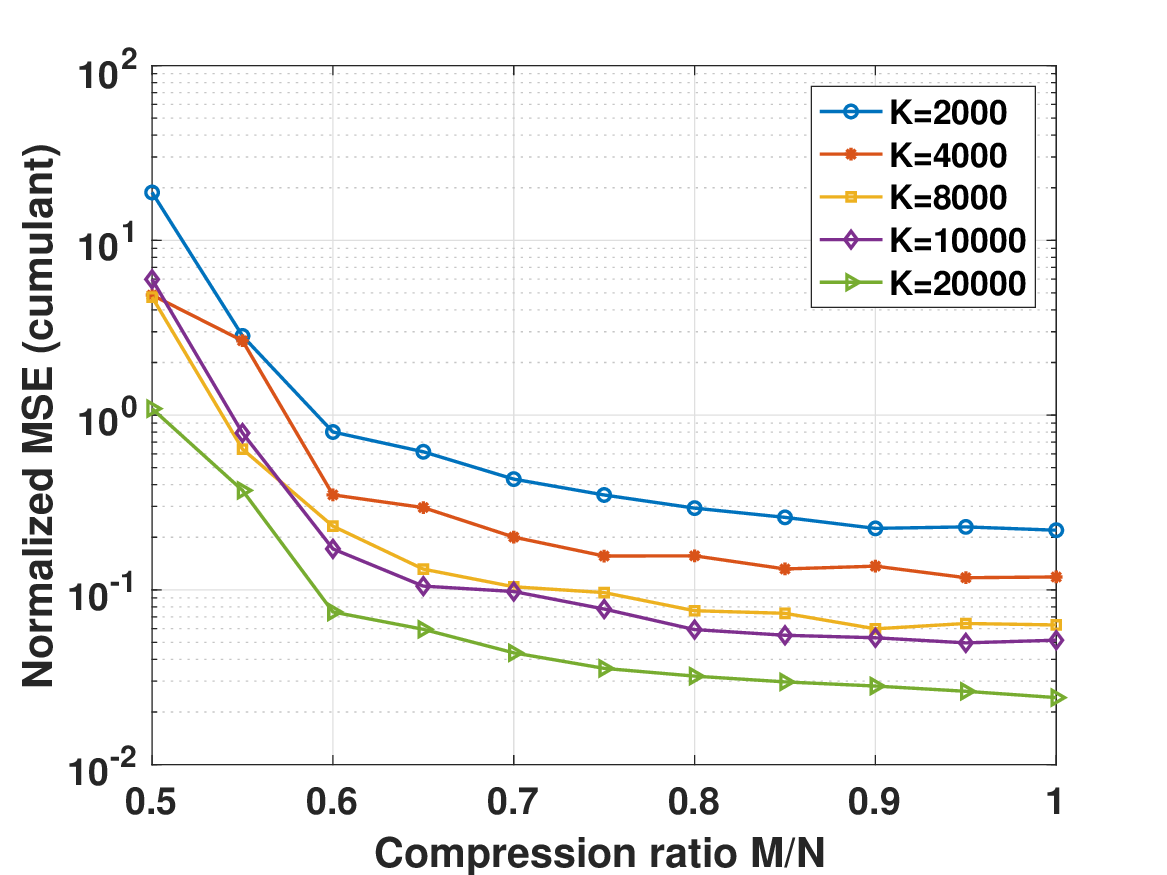}}
  \caption{\label{compression rate for cumulant estimation (noise-free)}MSE v.s. compression ratio for cumulant estimation (noise-free).} 
\end{figure}
\begin{figure}
  \centering
  \centerline{\includegraphics[width=9cm, height=5.5cm]{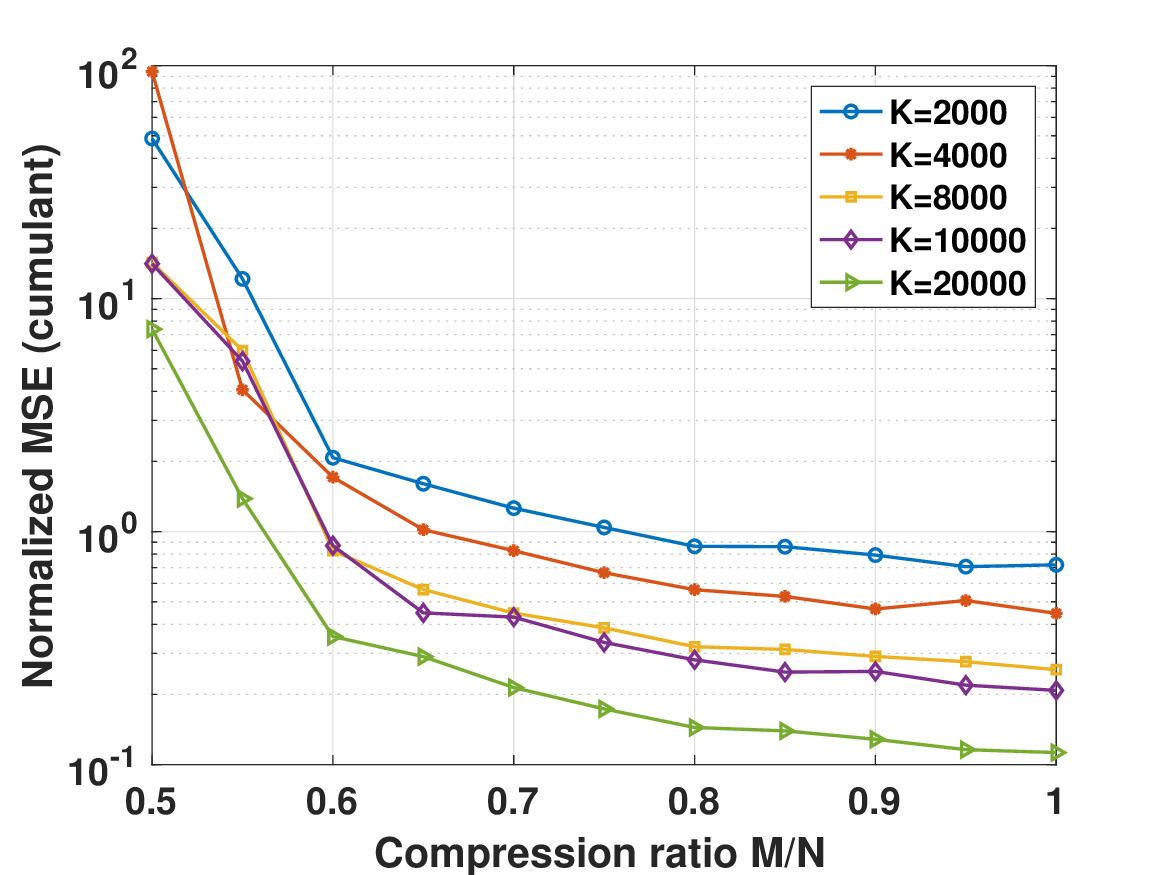}}
  \caption{\label{compression rate for cumulant estimation (colored Gaussian noise)}MSE v.s. compression ratio for cumulant estimation (colored Gaussian noise SNR = $0$ dB).} 
\end{figure}

First, it is of great interest to study the impact of the compression ratio on the error performance, we define the performance metric as the normalized mean-square error (MSE) of the estimated $\hat{\bar{\mathbf{c}}}_{3,\mathbf{x}}$ with respect to the uncompressed one, for a given compression ratio $M/N$. That is, $\rm{MSE}=E\left\lbrace\norm{\hat{\bar{\mathbf{c}}}_{3,\mathbf{x}}-\bar{\mathbf{c}}_{3,\mathbf{x}}}_2^2/\norm{\bar{\mathbf{c}}_{3,\mathbf{x}}}_2^2\right\rbrace$. In the \emph{noise-free setting}, Fig. \ref{compression rate for cumulant estimation (noise-free)} depicts the MSE of C3CS versus the compression ratio $M/N$, for a varying number of measurement vectors $K$. Fig. \ref{compression rate for cumulant estimation (noise-free)} shows that the estimation performance of C3CS is satisfactory even at a compression ratio $M/N=0.6$ without posing any sparsity constraint on the signal, and the performance improves with increasing $M/N$. This is understandable because weaker compression always leads to better performance, and $M/N=1$ stands for no compression. In this simulation, the sharp drop of MSE happens when $M/N=0.6$, so we say $0.6$ is the strongest achievable compression ratio. It is worth pointing out that the achievable strongest compression ratio $M/N$ can be much smaller than $0.6$ and even below $0.2$ when $N$ is large as illustrated in Fig. \ref{fig:  strongest compression}. In this simulation, $N$ is set to be small for a simple illustration. Furthermore, the error performance also improves with increasing $K$ but the improvement becomes slight when $K$ is large enough (observe the curves for $K=8000$ and $K=10000$ in Fig.  \ref{compression rate for cumulant estimation (noise-free)}). This is because a large number of data blocks can effectively reduce the finite sample effect and when $K$ is large enough the finite sample effect has been adequately alleviated.

Suppressing colored Gaussian noise is one of the most important merits of third-order cumulants. The error performance of C3CS in the \emph{colored Gaussian noise setting} is evaluated in Fig. \ref{compression rate for cumulant estimation (colored Gaussian noise)}, where the added  $0$ dB colored Gaussian noise is generated by passing white Gaussian noise through the MA(5) model with coefficients [1, -2.33, 0.75, 0.5, -1.3, -1.4]. In this setting, the estimation error comes from not only the compression but also the strong noise. Theoretically, the third-order cumulants of colored Gaussian noise are identical to zero. However, practical estimation only accesses to a finite number of data records, which can not average the third-order cumulants of colored Gaussian noise to be exactly zero. As a result, the performance of C3CS in Fig. \ref{compression rate for cumulant estimation (colored Gaussian noise)} is not as good as that in Fig. \ref{compression rate for cumulant estimation (noise-free)}. Even so, we can see that the third-order cumulants can still be recovered satisfactorily from compressive measurements as long as given a large number of data records ($K=8000$ or larger) such that the Gaussian noise is averaged to be nearly zero. The error performance also improves with increasing compression ratio $M/N$, which is similar to the trends in the noise-free setting. Fig. \ref{compression rate for cumulant estimation (colored Gaussian noise)} demonstrates the effectiveness of C3CS on the robustness to colored Gaussian noise, which is very useful in practical application as illustrated next.

\subsection{C3CS: Application to Correlation Function Estimation}
\label{ssec:subhead}

\begin{figure}
  \centering
  \centerline{\includegraphics[width=9cm, height=5.5cm]{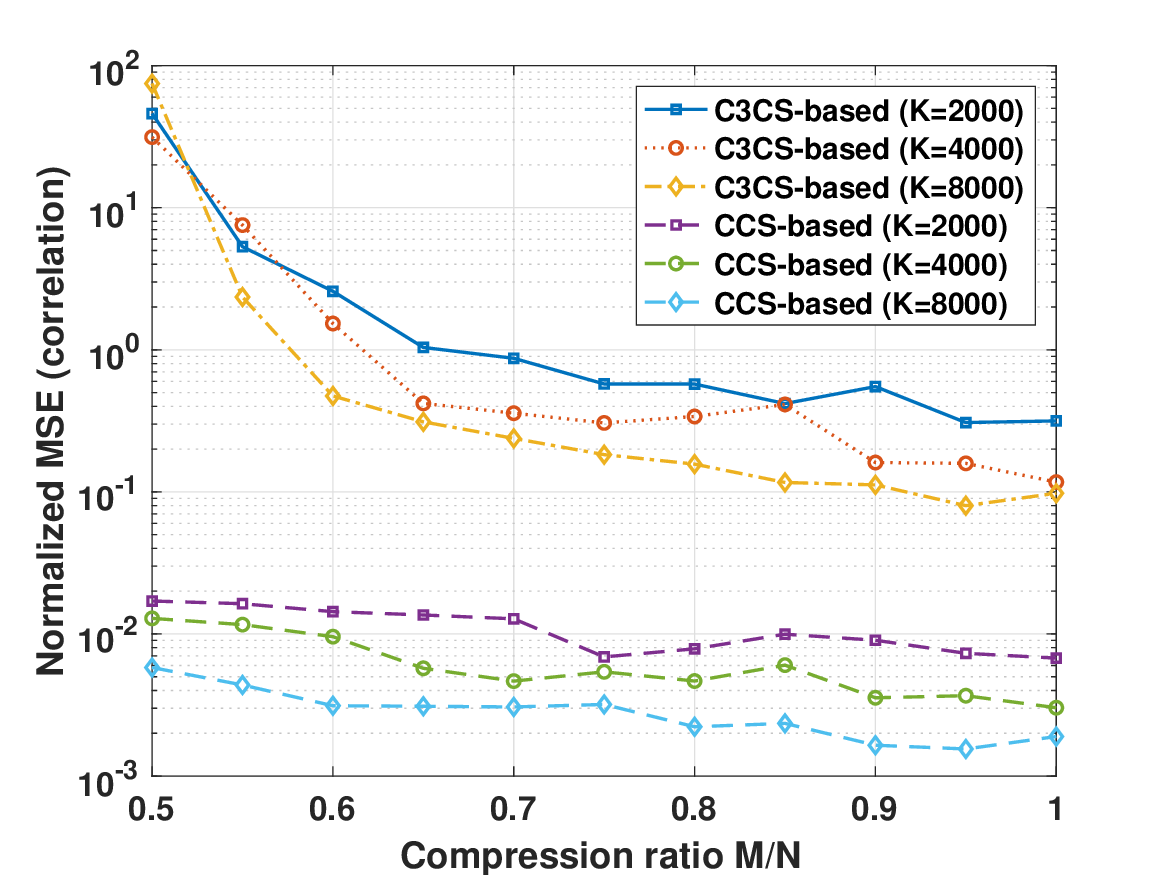}}
  \caption{\label{compression rate for correlation estimation based on C3CS and CCS (noise-free)}MSE v.s. compression ratio for correlation estimation based on C3CS and CCS (noise-free).} 
\end{figure}
\begin{figure}
  \centering
  \centerline{\includegraphics[width=9cm, height=5.5cm]{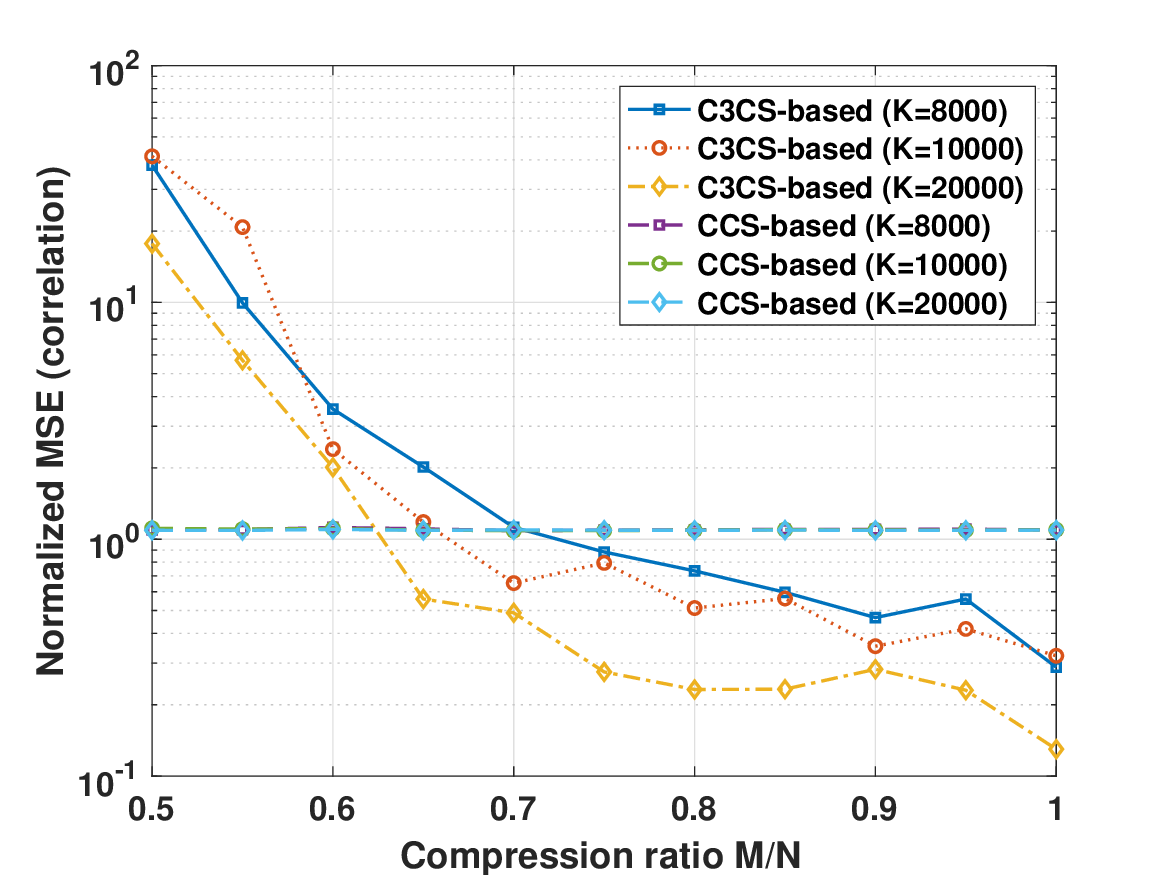}}
  \caption{\label{compression rate for correlation estimation based on C3CS and CCS (colored Gaussian noise)}MSE v.s. compression ratio for correlation estimation based on C3CS and CCS (colored Gaussian noise SNR = $0$ dB).} 
\end{figure}

The correlation function is useful in many signal processing tasks, such as spectral density estimation, DoA esitmation, and etc. However, under the strong colored Gaussian noise setting, the correlation estimate is highly biased if the spectrum of noise is unknown for prewhitening. In this section, we consider estimating the unbiased correlation function from compressive measurements by using C3CS even when the uncompressed signal is contaminated by strong colored Gaussian noise. Specifically, We estimate the unbiased correlation function based on the output of C3CS followed by the method proposed in \cite{giannakis1990nonparametric}. As a benchmark, we also evaluate the correlation function estimation based on CCS \cite{tian2011cyclic}. The data used for the simulation here is the same as in Part A.

In the \emph{noise-free setting}, the error performance of correlation estimation based on C3CS and CCS is depicted in Fig. \ref{compression rate for correlation estimation based on C3CS and CCS (noise-free)}. It signifies that CCS-based approach achieves better MSE performance than the C3CS-based one. This is because given the same number of data blocks, the variance of the covariance estimate is smaller than that of the cumulant estimate. Moreover, the strongest achievable compression ratio of C3CS is about $0.6$ in this simulation, while that of CCS is much smaller at least below $0.5$. In other words, CCS achieves stronger compression than C3CS as analyzed in Remark 4. For correlation function estimation application, although CCS based approach shows appealing advantages in the noise-free setting, we will show next how C3CS-based approach outperforms the CCS-based one in the colored Gaussian noise setting.

\begin{figure}
  \centering
  \centerline{\includegraphics[width=9cm, height=5.5cm]{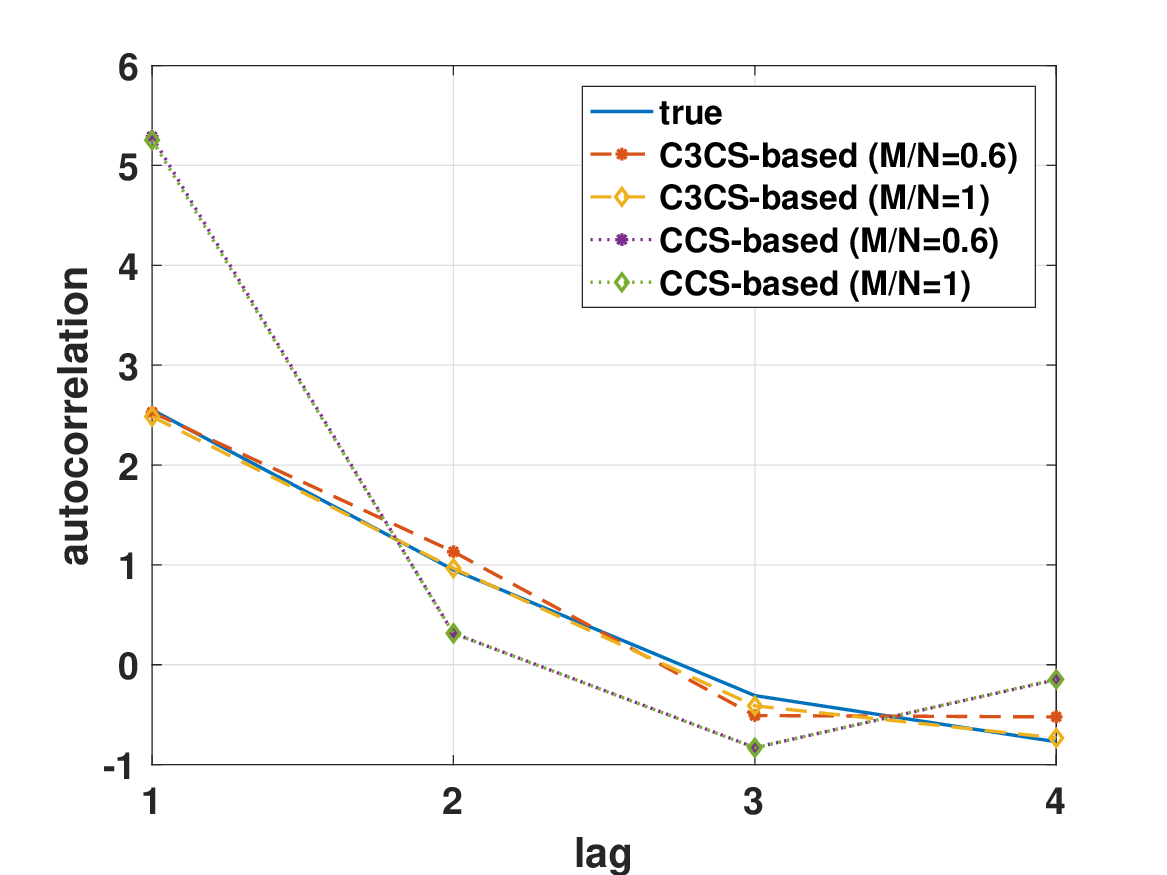}}
  \caption{\label{The estimated correlation values based on C3CS and CCS}The estimated correlation values based on C3CS and CCS ($K = 4000$,  colored Gaussian noise SNR = $0$ dB).} 
\end{figure}

In the \emph{colored Gaussian noise setting}, the error performance of correlation estimation based on C3CS and CCS is shown in Fig. \ref{compression rate for correlation estimation based on C3CS and CCS (colored Gaussian noise)}. It shows that the curves corresponding to CCS approach with various $K$ are almost overlapped and the corresponding error is always greater than $1$, which means that the CCS-based approach fails to estimate the correlation function (no matter how large $K$ is) due to the presence of strong colored Gaussian noise. In contrast, the C3CS-based approach still works as long as given enough data blocks to reduce the finite sampling effect and average the colored Gaussian noise to be nearly zero. We emphasize again that the achievable strongest compression ratio of C3CS can be much smaller when $N$ is large in practical applications. 

To gain a more straightforward comparison between the C3CS-based and CCS-based approaches for correlation estimation, we also plot the estimated correlation values at several lags under different compression ratios in Fig. \ref{The estimated correlation values based on C3CS and CCS}, while the true correlation values are also plotted in the solid line as a benchmark. The estimated correlation values obtained from both the C3CS-based and the CCS-based approaches are calculated by averaging over multiple Monte Carlo runs. We can see that the CCS-based approach is highly biased even without any compression ($M/N=1$) while the C3CS-based approach achieves quite good estimation even with $M/N=0.6$, which validates the superiority of the proposed C3CS approach in terms of sampling rate reduction and noise suppression.

\subsection{CCSS: Robustness to Sampling Rate Reduction}
\label{ssec:subhead}

In this subsection, we consider a signal $x(t)$ containing $2$  harmonics with analog frequencies $f_1=1MHz$ and $f_2=2MHz$ and phases are i.i.d. uniformly distributed over [-$\pi$, $\pi$]. Assume a conservative sampling rate $f_s=10MHz$ to get a digital counter part $x(n)$ with digital frequencies $w_1=0.1$ and $w_2=0.2$. It has been shown in \cite{swami1991cumulant} that both cumulant slice $c_{4,x}(\tau, \tau, \tau)$ and correlation $c_{2,x}(\tau)$ can be used to retrieve harmonics contained in $x(t)$, while the former is more robust to colored Gaussian noise. In this simulation, $c_{4,x}(\tau, \tau, \tau)$ and $c_{2,x}(\tau)$ are estimated from $K$ independent realizations, each consisting of $N=16$ samples.  We sample $x(n)$ at the sub-Nyquist rate via (\ref{yphix}) to test CCSS.

\begin{figure}
  \centering
  \centerline{\includegraphics[width=9cm, height=5.5cm]{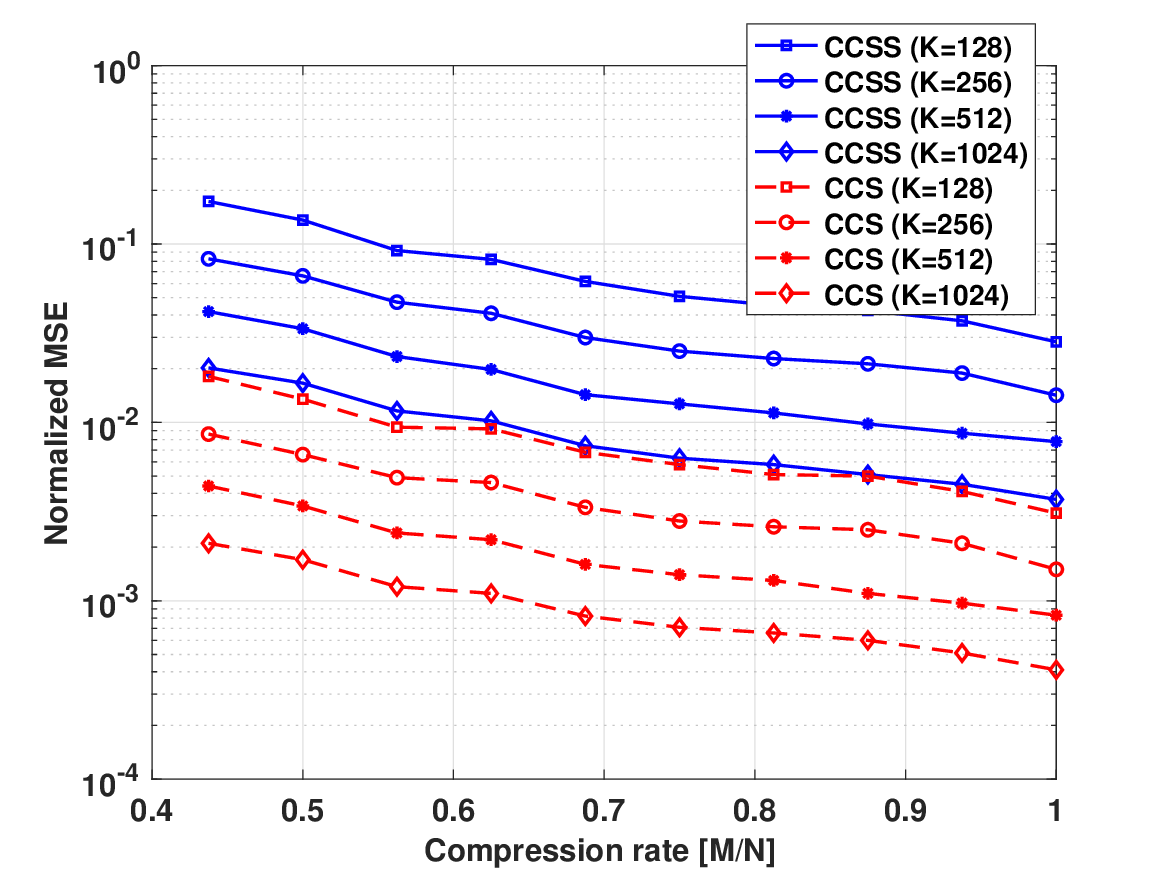}}
  \caption{\label{compression rate for CCSS and CCS over various number of data blocks (noise-free)}NMSE v.s. compression rate for CCSS and CCS over various number of data blocks (noise-free).} 
\end{figure}
\begin{figure}
  \centering
  \centerline{\includegraphics[width=9cm, height=5.5cm]{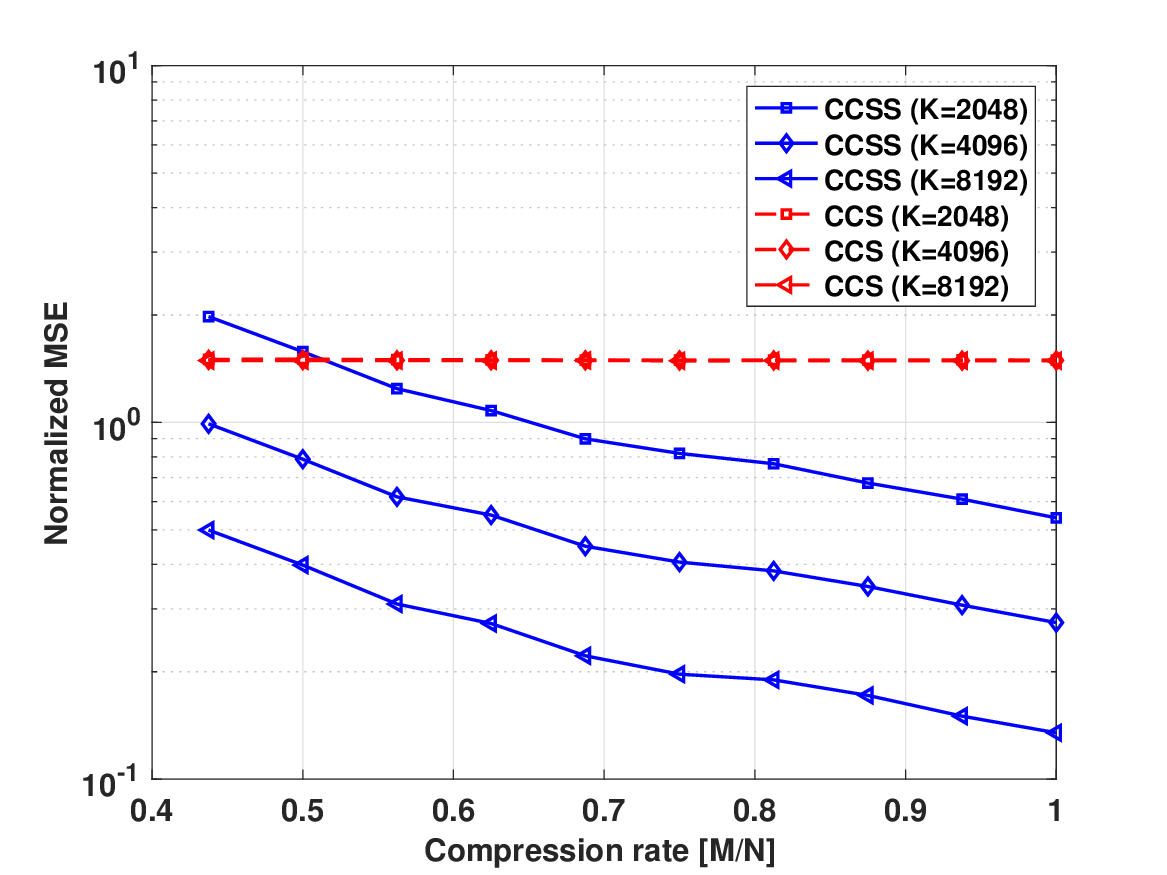}}
  \caption{\label{compression rate for CCSS and CCS over various number of data blocks (colored Gaussian noise)}NMSE v.s. compression rate for CCSS and CCS over various number of data blocks (colored Gaussian noise SNR = $0$ dB).} 
\end{figure}

We first examine the robustness to the sampling rate reduction of our proposed CCSS approach. Since the length-16 ($N=16 $) minimal sparse ruler has $K=7$ distance marks, the sampler $\boldsymbol{\Phi}$ is designed by choosing the 7 rows corresponding to the distance marks from the identity matrix $\mathbf{I}_{16}$. In this case, the sampler achieves the strongest compression rate $M/N \approx 0.43$. Then the larger $M/N$ cases are generated by randomly adding additional rows of $\mathbf{I}_{16}$ to the already chosen 7 rows. The normalized MSE between the estimated cumulant slice and the true one is calculated for sparse ruler sampling ($K<N$).  In the \emph{noise-free setting}, Fig. \ref{compression rate for CCSS and CCS over various number of data blocks (noise-free)} shows the MSE under various values of $K$. It indicates that the estimation performance of CCSS is satisfactory even when $M/N \approx 0.43$, and can be improved with increasing $M/N$ and $K$. In addition, Fig. \ref{compression rate for CCSS and CCS over various number of data blocks (noise-free)} also shows that CCS achieves better MSE performance than CCSS for fixed $M/N$ and $K$, which is due to the faster convergence rate of sample second-order statistics. However, the superiority of CCSS arises when strong colored Gaussian noise ($0$ dB) is present as shown in Fig. \ref{compression rate for CCSS and CCS over various number of data blocks (colored Gaussian noise)} where the added colored Gaussian noise is generated by passing white Gaussian noise through an ARMA filter with AR parameters [1, 1.4563, 0.81] and MA parameters [1, 2, 1]. It turns out that CCSS still works as long as given large enough $K$ to reduce finite sample effect and hence suppressing Gaussian noise, while, in contrast, CCS does not work at all because its estimation error is almost constant greater than $1$. We next show the advantage of C3CS in a practical task.

\subsection{CCSS: Application to Line Spectrum Estimation}
\label{ssec:subhead}

\begin{figure}
  \centering
  \centerline{\includegraphics[width=9cm, height=5.5cm]{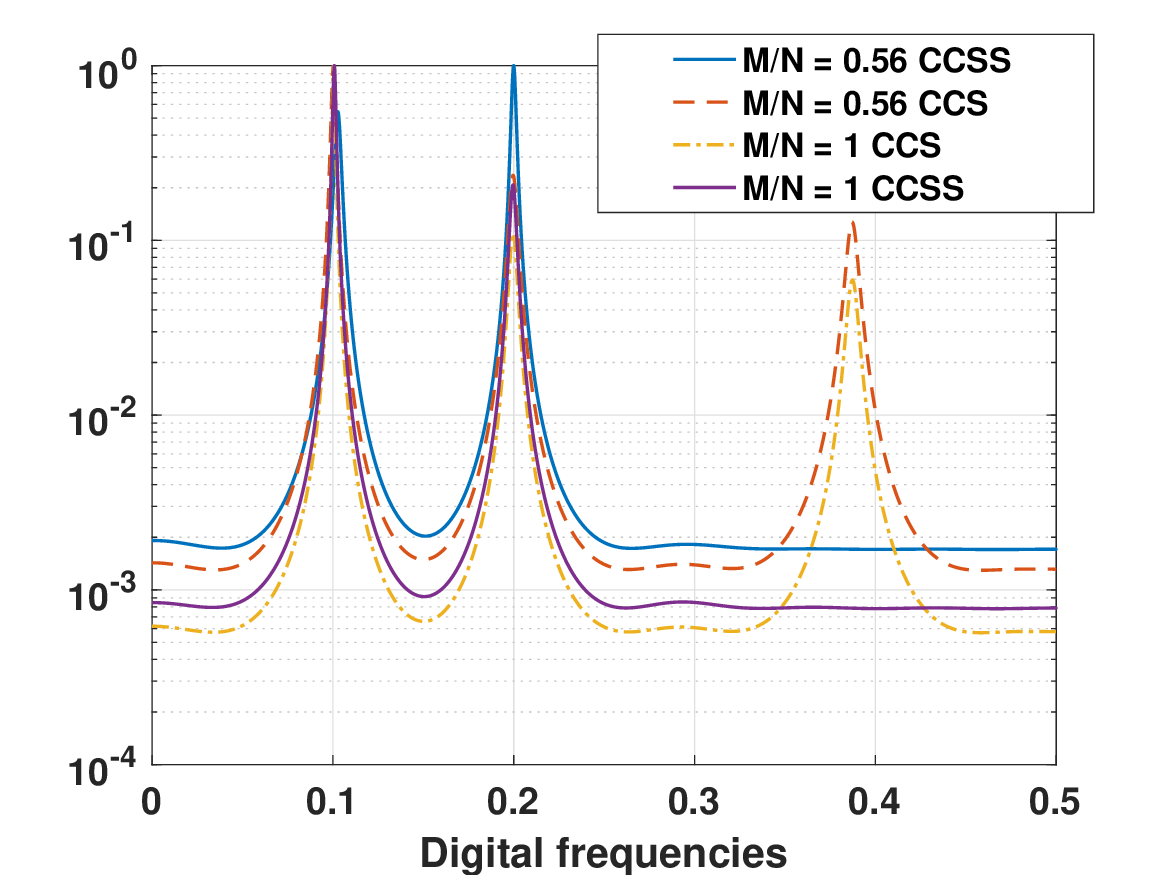}}
  \caption{\label{MUSIC spectrum based on the cumulant slice and the covariance}MUSIC spectrum based on the cumulant slice and the covariance with different compression rate [M/N]  (K = 4096, colored Gaussian noise SNR = $0$ dB).}     
\end{figure}

In this subsection, we test the performance of CCSS in the line spectrum estimation application and the data used here is the same as in Part C. Specifically, we first estimate the cumulant slice $c_{4,x}(\tau, \tau, \tau)$ and the correlation $c_{2,x}(\tau)$ based on CCSS and CCS respectively, and then use the estimates to retrieve harmonics as in \cite{swami1991cumulant}.

Fig. \ref{MUSIC spectrum based on the cumulant slice and the covariance} depicts the MUSIC spectrum based on CCS and CCSS. Here, the cumulant slice estimate and the covariance estimate are calculated from $K=4096$ data blocks under the same noise setting as above. Note this colored noise spectrum has a strong pole at frequency $w=0.4$. We can see that the frequency estimation performance is already quite good when $M/N=0.56$ for both the cumulant slice and covariance based methods, and can be further improved by increasing the compression rate to $M/N=1$. In practice, the choice of the compression rate $M/N$ reflects a trade-off between estimation accuracy and sampling rates. Further, the MUSIC spectrum based on CCSS detects two peaks at $w_1=0.1$ and $w_2=0.2$, while that based on CCS shows an additional peak around $w=0.4$ resulted from the colored noise. Evidently, the estimated spectrum based on CCSS outperforms that of CCS in terms of removing colored Gaussian noise, which is particularly useful when the noise spectrum is unavailable or hard to estimate for prewhitening purpose.

\section{Summary}
This paper presents novel approaches to the reconstruction of third-order statistics from compressive measurements. Capitalizing on the structures of third-order cumulants resulting from signal stationarity and cumulant symmetry, an over-determined linear system is constructed that explicitly links the cross-cumulants of compressive samples with the cumulants of the uncompressed signal. Guideline on compressive sampler design is provided, and the conditions for lossless recovery guarantee are delineated. It is proven that the proposed techniques are able to achieve significant compression, even when no sparsity constraint is imposed on the signal or cumulants. For applications where only the diagonal cumulant slice is required, a computationally simple approach is developed to directly reconstruct the desired cumulant slices from compressive samples without having to recover the entire cumulant tensor, which turns out to subsume the existing compressive covariance sensing framework as a special case. Extensive simulations are conducted corroborating the effectiveness of the proposed techniques.



\appendices

\section{Mapping Matrix $\mathbf{P}_N$}
This is to show how the mapping matrix $\mathbf{P}_N$ in (\ref{PN}) is determined. Note that $\tilde{\mathbf{c}}_{3,\mathbf{x}} \in \mathbb{R}^{N(N+1)/2 \times 1}$ and $\bar{\mathbf{c}}_{3,\mathbf{x}}\in \mathbb{R}^{N^3\times 1}$ contain the same set of unique cumulant values, and thus the mapping matrix we aim at finding out is actually a tall deterministic binary matrix $\mathbf{P}_N  \in \left\lbrace0, 1\right\rbrace^{N^3 \times N(N+1)/2}$. We first introduce a basic fact that will facilitate our derivation: Denote the $(i_1, i_2, i_3)$th entry of the cumulant tensor $\mathbfcal{C}_{3,x}$ as $\mathbfcal{C}_{3,x}(i_1,i_2,i_3)=c_{3,x}(i_2-i_1,i_3-i_1)$, then after vectorization this entry is mapped into the vector $\bar{\mathbf{c}}_{3,\mathbf{x}}$ with index being $(i_1-1)N^2+(i_2-1)N+i_3)$, say $\bar{\mathbf{c}}_{3,\mathbf{x}}((i_1-1)N^2+(i_2-1)N+i_3)=c_{3,x}(i_2-i_1,i_3-i_1)$. 

Recall that $\tilde{\mathbf{c}}_{3,\mathbf{x}}$ is defined as
\begin{equation}
\begin{aligned}
\tilde{\mathbf{c}}_{3,\mathbf{x}} = &[c_{3,x}(0, 0), c_{3,x}(0, 1), \cdots, c_{3,x}(0, N-1),\\
&c_{3,x}(1, 1), c_{3,x}(1, 2), \cdots, c_{3,x}(1, N-1),\\
&\cdots \cdots, c_{3,x}(N-1, N-1)]^T.
\end{aligned} 
\end{equation}
Observe that $c_{3,x}(v, u), \forall u\in[0, N-1], v\in[u,N-1]$, the elements in $\tilde{\mathbf{c}}_{3,\mathbf{x}}$, can be found in $\mathbfcal{C}_{3,x}$ at indices $(w, w+v, w+u),  \forall u\in[0, N-1], v\in[u,N-1], w\in[1,N-v]$ and indices $(w, w+u, w+v), (w+v, w, w+u, ), (w+v, w+u, w), (w+u, w, w+v), (w+u, w+v, w)$ due to the symmetry Property 1. Then equivalently, according to the fact introduced above, $c_{3,x}(v, u), \forall u\in[0, N-1], v\in[u,N-1]$ can also be found in $\bar{\mathbf{c}}_{3,\mathbf{x}}$ at indices
\begin{equation}
\begin{aligned}
&p_1(u,v,w) = (w-1)N^2+(w+v-1)N+(w+u)\\
&p_2(u,v,w) = (w-1)N^2+(w+u-1)N+(w+v)\\
&p_3(u,v,w) = (w+v-1)N^2+(w-1)N+(w+u)\\
&p_4(u,v,w) = (w+v-1)N^2+(w+u-1)N+w\\
&p_5(u,v,w) = (w+u-1)N^2+(w-1)N+(w+v)\\
&p_6(u,v,w) = (w+u-1)N^2+(w+v-1)N+w\\
&\forall u\in [0, N-1], v\in [u, N-1], w \in [1, N-v].
\end{aligned}
\end{equation}
Concisely, we have
\begin{equation}
\bar{\mathbf{c}}_{3,\mathbf{x}}(p_i(u,v,w))=c_{3,x}(v, u), \quad i=1,\cdots,6, \label{puv}
\end{equation}
for $\forall u\in [0, N-1], v\in [u, N-1], w \in [1, N-v]$.

Next we need to find out the indices of $c_{3,x}(v, u), \forall u\in[0, N-1], v\in[u,N-1]$ located in $\tilde{\mathbf{c}}_{3,\mathbf{x}}$. It is easy to validate that $\tilde{\mathbf{c}}_{3,\mathbf{x}}(v-u+1)=c_{3,x}(v, u),  \forall u=0, v\in[u,N-1]$ and $\tilde{\mathbf{c}}_{3,\mathbf{x}}(\sum_{j=1}^u(N-j+1)+v-u+1)=c_{3,x}(v, u), \forall u\in[1,N-1], v\in[u,N-1]$, or equivalently we say the indices of $c_{3,x}(v, u), \forall u\in[0, N-1], v\in[u,N-1]$ in $\tilde{\mathbf{c}}_{3,\mathbf{x}}$ are
\begin{equation}
\begin{aligned}
&q(u,v) = \begin{cases} 
      v-u+1 & u=0\\
      \sum_{j=1}^u(N-j+1)+v-u+1 & u\neq 0
   \end{cases}\\
&\forall u\in [0, N-1], v\in [u, N-1].
\end{aligned} 
\end{equation}
Concisely, we have
\begin{equation}
\tilde{\mathbf{c}}_{3,\mathbf{x}}(q(u,v) )=c_{3,x}(v, u), \label{quv}
\end{equation}
for $\forall u\in [0, N-1], v\in [u, N-1]$.

To complete, we conclude from (\ref{puv}) and (\ref{quv}) that $\tilde{\mathbf{c}}_{3,\mathbf{x}}(q(u,v))= \bar{\mathbf{c}}_{3,\mathbf{x}}(p_i(u,v,w)),  i=1,\cdots,6$ for $\forall u\in [0, N-1], v\in [u, N-1], w \in [1, N-v]$. Combining with the relationship $\bar{\mathbf{c}}_{3,\mathbf{x}} = \mathbf{P}_N\tilde{\mathbf{c}}_{3,\mathbf{x}}$, we determine the mapping matrix $\mathbf{P}_N$ as
\begin{equation}
\begin{cases}
\left[\mathbf{P}_N\right]_{\left(p_i(u,v,w), q(u,v)\right)} = 1, \quad  i = 1,2,...,6,\\
\left[\mathbf{P}_N\right]_{otherwise} = 0,
\end{cases}\\ 
\end{equation}

%

\ifCLASSOPTIONcaptionsoff
  \newpage
\fi



%

\bibliographystyle{IEEEbib}
\bibliography{strings,refs}

\end{document}